\documentclass[journal,twoside]{IEEEtran}
\usepackage{cite}
\usepackage{amsmath,amssymb,amsfonts}

\usepackage{algorithmic}
\usepackage{graphicx}
\usepackage{textcomp}
\usepackage{subfigure}
\usepackage{booktabs}
\usepackage{makecell}
\usepackage{multirow}
\usepackage{setspace}
\usepackage[table]{xcolor}

\usepackage{longtable,rotating}
\usepackage{threeparttable}

\usepackage[numbers]{natbib}
\usepackage{color}

\def\BibTeX{{\rm B\kern-.05em{\sc i\kern-.025em b}\kern-.08em
    k\kern-.1667em\lower.7ex\hbox{E}\kern-.125emX}}
\begin{document}

\title{SOC-Boundary and Battery Aging Aware Hierarchical Coordination of Multiple EV Aggregates Among Multi-stakeholders with  Multi-Agent Constrained Deep Reinforcement Learning}
\author{Xin~Chen*
   
\thanks{This work was supported in part by the National Natural Science Foundation of China (Grant No.21773182 (B030103)) and the HPC Platform, Xi'an Jiaotong University.}
\thanks{Xin Chen (corresponding authour, e-mail: xin.chen.nj@xjtu.edu.cn) is with Center of Nanomaterials for Renewable Energy, State Key Laboratory of Electrical Insulation and Power Equipment, School of Electrical Engineering, Xi'an Jiaotong University, Xi'an, Shaanxi, China.}
}

\maketitle

\begin{abstract}

As electric vehicles (EV) become more prevalent and advances in electric vehicle electronics continue, vehicle-to-grid (V2G) techniques and large-scale scheduling strategies are increasingly important to promote renewable energy utilization and enhance the stability of the power grid. This study proposes a hierarchical multistakeholder V2G coordination strategy based on safe multi-agent constrained deep reinforcement learning (MCDRL) and the Proof-of-Stake algorithm to optimize benefits for all stakeholders, including the distribution system operator (DSO), electric vehicle aggregators (EVAs) and EV users. For DSO, the strategy addresses load fluctuations and the integration of renewable energy. For EVAs, energy constraints and charging costs are considered. The three critical parameters of battery conditioning, state of charge (SOC), state of power (SOP), and state of health (SOH), are crucial to the participation of EVs in V2G. Hierarchical multi-stakeholder V2G coordination significantly enhances the integration of renewable energy, mitigates load fluctuations, meets the energy demands of the EVAs, and reduces charging costs and battery degradation simultaneously.

\end{abstract}


\section{Introduction}
\label{sec:introduction}
\subsection{Background and Motivation}
 
Electric vehicles (EVs) are gaining traction due to several merits of price reduction and climate and environmental awareness, which neither emit tailpipe pollutants $NO_2$ nor $CO_2$ and have lower maintenance cost and energy cost \cite{sanguesa2021review}. Renewable energy sources (RES), such as wind and solar, can provide energy for electric vehicles without greenhouse gases. Due to the intermittency and unpredictability of renewable energy, energy storage systems should be combined to improve the power quality of renewable energy. Vehicle-to-grid (V2G) is a widely recognized solution in which EVs play an important role as a storage resource for the power grid\cite{bibak2021comprehensive}. 
Despite the advantages of electric vehicles, the significant load pressure and randomness of charging behaviors could challenge the stability and economic effectiveness of distribution system operators (DSOs). 
\paragraph{DSO Coordination Strategy}
Studies have been carried out to address the potential problems mentioned above, which could be divided into two categories: market-based and control-based\cite{jin2020local}.
Both of them aim to alleviate congestion and operate in a more economically efficient way through flattening out the load curves \cite{kabiri2024transactive,crozier2020opportunity,singh2020cost}. 
The former, market-based stream utilizes the market tools, such as marginal prices that distribution system operators publish in advance to motivate EV aggregators (EVA) to adjust their demand for lower charging costs\cite{kabiri2024transactive}. Mutual benefit and reciprocity could be achieved by this economic means. Therefore making a reasonable prediction for the wave of EVA energy demand and pricing strategy is crucial for maximizing profits. 
Shinde et al.\cite{shinde2022modified} modeled the participation of an EVA in the intraday and balancing markets as a multistage stochastic programming problem, with the purpose of reducing the computational complexity due to peculiarities of the intraday market. 
Liu et al.\cite{liu2020development} proposed a robust optimization model for EVAs with varying risk preferences and suggested a step-wise bidding strategy based on optimized schedules to account for potential market price fluctuations.  

By comparison, the control-based stream concentrates on the direct optimization of charging/discharging scheduling,  with its effectiveness quantified by Crozier et al \cite{crozier2020opportunity}  and some combined this active power dispatch with reactive power injection \cite{singh2020cost}. 
Notably, this paper mainly focuses on the control-based method using V2G technology.
As a type of distributed energy storage technology, V2G has been gaining increasing attention for its potential to improve EV utilization rates and charging/discharging piles\cite{xu2022short}, and regulate frequency\cite{abubakr2023novel}, voltage\cite{hu2021distributed}, and peak shaving\cite{zhang2022transfer} through the implementation of appropriate scheduling strategies. 
In the V2G coordination strategies, the players, including EVs and power grids, are complex, which brings difficulties and security issues. 
In the intelligent contracts of the V2G trading, two-level optimization strategies were constructed to achieve a fair distribution of benefits among the different market participants, namely EVs, electric vehicle aggregators (EVAs), and the power grid \cite{luo2022hierarchical}. Information sharing between different participants and user privacy protection issues is an important question. For this reason, Luo et al. proposed a vehicle-to-vehicle and V2G electricity trading architecture based on blockchain\cite{luo2021blockchain}. 

Deep reinforcement learning (DRL), whether applied to single-agent or multi-agent scenarios, exhibits superior ability to solve unsupervised, nonlinear, and high-dimensional problems in power system compared to other optimization algorithms\cite{qiu2023reinforcement}.
A DRL technique based on Deep Q-Networks (DQN) is presented to determine the best charging strategy for EVs, considering the heterogeneity of empirical travel patterns and the unpredictability of electricity prices\cite{hao2023v2g}. As for continuous action control, the proximal policy optimization (PPO) algorithm can provide more precise control\cite{schulman2017proximal}. 

There has been growing interest in integration of multiple parties including power generation, distribution and demand to improve energy efficiency. Particularly, Multi-agent DRL has shown a considerable advantage over existing physical models in terms of its ability to deal with problems with high complexity\cite{perera2021applications}. Fang et al. proposed a load frequency control strategy for multi-microgrids with V2G based on an enhanced Multi-agent Deep Deterministic Policy Gradient (MADDPG) \cite{fan2023load}. 
Considering the  satisfaction of EV owners, a decentralized MADRL framework for EV charging was introduced, which aimed to optimize the loss of life (LOL) of distribution transformer\cite{li2022multiagent, li2022ev}. However, the number of EVs is too small in their work, and the aging batteries are poorly modeled.
Wang et al. presented a novel real-time parameterized double DQN approach to accommodate uncertainties in RESs, load profiles, line outages, and traffic volumes, addressing the resilience-driven routing and scheduling problem of mobile energy storage systems in power and transportation networks after extreme events\cite{wang2022multi}. However, the algorithm is discrete and does not have the capability to manage the continuous charging/discharging power of large-scale EVs.  
A novel improved MADDPG framework tackles the allocation of EV charging station by optimizing EV continuous charging schedules while evaluating technical (power loss, voltage stability), economic (installation, operating expenses), and environmental (carbon emissions) objectives\cite{abid2024novel, qiu2021scalable}. 
The counterfactual multi-agent policy gradients algorithm was used to efficiently charge multiple EVs to minimize costs in a smart grid that includes a PV and energy storage system\cite{park2022multi}.  
To address the challenges of privacy considerations, innovative model-free MADRL approaches for the decentralized coordination of energy hubs and EVAs in integrated energy distribution and harvesting systems were introduced\cite{zhang2023novel, zhang2023distributed}. 
The VDAC-BLP algorithm proposed effectively handles dynamic constraints in the joint scheduling of trucks and battery charging stations in battery swapping-charging systems for EVs in urban settings, demonstrating improved computational efficiency and performance\cite{liang2022real}.


Considering the widely connected nature of the power grid and the potential instability issues, even small violations of physical constraints are not allowed in the DSO scheduling process\cite{qiu2023reinforcement}.
Hence, safer and more robust agents are expected to be studied, whose testing process needs to ensure zero constraint violations.
While numerous studies have focused on restricting agent actions by adding penalty terms that measure the danger level of decisions to the reward function and fine-tune their performance\cite{tuchnitz2021development,ding2020optimal}, the design of effective and comprehensive penalty coefficients has proven to be challenging. Furthermore, the adjustment of the policy often lags behind the travel System-on-Chip (SoC) requirements, making it challenging to give a feasible solution using the data-driven RL method considering the computation cost.

Lacking a mechanism to deal with constraints in the physical world seems to be an inherent flaw in previous RL algorithm, until some breakthroughs, such as the accelerated primal-dual policy optimization incorporating off-policy training \cite{liang2018accelerated} and Constrained Policy Optimization (CPO) \cite{achiam2017constrained}. Great contributions have been made on the application of Safe RL in DSO management since then. 
Gao et al. developed a practical Centralized Training, Decentralized Execution (CTDE) platform, where the long-term cost constraints are tackled through primal-dual optimization and can automatically adjust the penalty term\cite{gao2023cloud}. However, they mainly focus on agents in discrete action space, which is incompatible with ours. Furthermore, in \cite{li2019constrained}, the EV charging schedule problem highlighting users' demand for fully charged state was formulated as a Constrained Markov Decision Process (CMDP) and solved by policy optimization using Lagrange relaxation technique.
However, their paper failed to show how to generalize their method to large-scale agents, a crucial consideration for the non-stationary training process in dynamic settings.
Since single agent cannot accurately depict the real scenario, we have further developed this approach and integrated it into a multi-agent setting.
Another straightforward approach for dealing with problems involving constrained feasible regions is to incorporate a stand-alone safety module that will activate whenever a violation occurs\cite{lee2023three}. Although their method did not solve the fundamental problem, it completely and effectively removes the unsafe tries. In this paper, we integrate the safeguard technique with a constrained policy optimization algorithm to completely eliminate violations.

\paragraph{Battery Aging Aware Coordination}

Battery aging is a significant factor that prevents electric vehicle customers from participating in V2G. Therefore, addressing the battery degradation issue in V2G scheduling is imperative. Wang et al. used a long-term and short-term memory neural network to capture long-term dependencies in the degraded capacities of lithium-ion (Li-ion) batteries\cite{wang2018dynamic}. Fang et al. conducted a deep analysis of battery degradation physics and developed an aging-effect coupling model based on an existing improved single-particle model\cite{fang2023performance}. Augello et al. described a blockchain-based approach for sharing data used both for monitoring the health of the EVs' batteries and for remuneration in V2G for tracking battery usage to permit second-life applications \cite{augello2023certifying}. Optimization models to optimize the charging/discharging schedule with V2G capabilities were proposed, aiming to minimize the charging costs while also considering battery aging, energy costs, and charging costs for the customers \cite{manzolli2022electric, ebrahimi2020stochastic,li2020optimization}. However, these strategies need to consider the complex relationship between battery aging and state of power (SOP) and state of charge (SOC), which deviates from the realistic operational needs of EVs.
\paragraph{EV Battery Simulation Model}
A battery is the central part of EVs participating in V2G. Different battery states SOX, such as SOC, SOP, and state of health (SOH), play various roles in the V2G strategy\cite{shrivastava2023review}. 
The SOC refers to the energy stored in an EV's battery at a given time. The SOP refers to the ability to provide power over a specific time under the design constraints of EVs' battery. Furthermore, due to concerns about potential damage to EV batteries from participating in V2G, participants can use SOH to measure the health of EV batteries. Therefore, building the SOX model considering battery SOC, SOP, and SOH is necessary. Accurate linear battery charging models were formulated, closely approximating the real-life battery power and SOC constraints \cite{pandvzic2018accurate, sakti2017enhanced}. In capturing the nonlinearities of Li-ion batteries, Naseri {\it{et al.}} proposed an efficient modeling approach based on the Wiener structure to reinforce the capacity of classical equivalent circuit models\cite{naseri2021enhanced}. For the characterization of the steady-state operation of Li-ion batteries, the model characterized battery performance, including non-linear charging/discharging power limits and efficiencies as a function of SOC and the requested power\cite{gonzalez2019non}.

\subsection{Scope and Contributions}

This article introduces a novel battery conditioning multi-stakeholder hierarchical V2G coordination strategy (MHVCS) based on Multi-Agent Constrained Policy Optimization (MACPO) and Proof of Stake (PoS) algorithms designed to achieve multi-stakeholder benefits. These benefits align with the interests of the distribution system operator (DSO), electric vehicle aggregators (EVA), and users. The strategy effectively schedules large-scale V2G operations by considering constraints at the EV, EVA, and power grid levels. Furthermore, at the lower level, the EVA power allocation strategy, based on the PoS algorithm, optimizes the power allocation to individual EVs. Integrating MACPO and PoS algorithms facilitates coordinating large-scale constrained V2G EV dispatching, addressing continuous charging/discharging challenges. The specific contributions of this article are as follows.
\begin{itemize}
\item First, the proposed MHVCS addresses the benefits for Distribution System Operators (DSO), Electric Vehicle Aggregators (EVA), and EV users, promoting advantages across all stakeholder groups.

\item Second, integrating the MACPO and Proof of Stake algorithms enables efficient large-scale V2G coordination. The approach accounts for fluctuations in grid load, renewable energy consumption, energy constraints, charging costs, and battery states of operation (SOX), including state of charge (SOC), state of power (SOP), and state of health (SOH).

\item Finally, the proposed MHVCS demonstrates its effectiveness in large-scale V2G coordination by considering the travel SOC boundary through the safe SOC constraint handling capabilities of the MACPO algorithm, a Multi-Agent Deep Reinforcement Learning approach. MHVCS can significantly enhance renewable energy consumption, mitigate load fluctuations, meet the energy demands of EVAs, reduce operating costs for DSOs, and minimize battery degradation.
\end{itemize}

\section{Problem Formulation}
\begin{figure*}[htbp]
\centering
\includegraphics[width=0.9\textwidth]{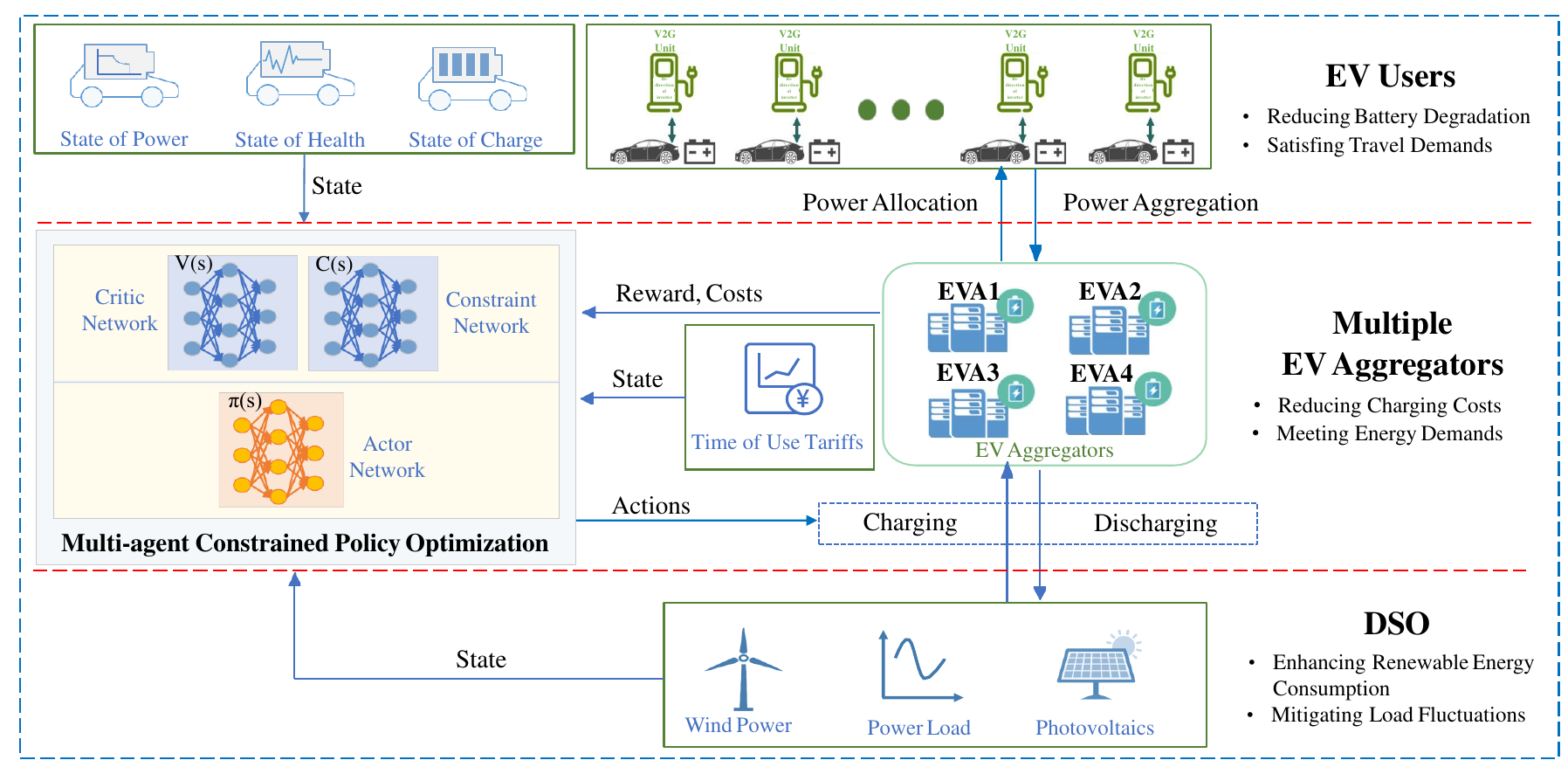}
\caption{The workflow of the proposed multi-stakeholder hierarchical V2G coordination strategy.}
\label{fig:macpoflow}
\end{figure*}

The workflow of MHVCS is presented in Fig.~\ref{fig:macpoflow}. The power system in the workflow comprises wind turbines (WTs), photovoltaics (PVs), electric vehicles (EVs), and distributed energy installations, which are collectively managed and controlled by multiple EVAs with MADRL. The three stakeholders are the EV user, EVA, and DSO.
The MADRL-based EVA coordination strategy can manage and optimize the battery SOX of the examined EVA as an intermediary entity between the EVs and the power grid. The problem is formulated as follows:
\begin{align}
&\min\limits_{P_{EVA}^{k}} F_1+ F_2+ F_3\label{pwgoal}\\
&F_1={\alpha}{\sigma^{2}}+\psi{f_1}\\
&F_2=\sum_{k=1}^Kc^{k}P_{EVA}^{k}\Delta k \label{F2}\\
&F_3=\sum_{n=1}^{N}{(c_{bat}Q_n+\frac{c_{l}}{1-SOH_{min}}){(1-SOH^i_n)}}\label{cbattery}
\end{align}
subject to:
\begin{align}
& \sum_{n \in \mathcal{N}} \max \left(\underline{e_n^{k_1}}-\overline{e_n^{k_2}}, \sum_{k_2+1}^{k_1} \underline{p_n^k} \Delta k\right) \nonumber\\
& \quad \leq E^{k_1}-E^{k_2} \leq \nonumber\\
&\sum_{n \in \mathcal{N}} \min \left(\overline{e_n^{k_1}}-\underline{e_n^{k_2}}, \sum_{k_2+1}^{k_1} \overline{p_n^{k}} \Delta k\right)\label{E}\\ 
&\underline{P_{grid}} \leq P^k \leq S*cos\phi\label{ptie}\\
&\underline{P_{EVA}^k} \leq P_{EVA}^k \leq \overline{P_{EVA}^k}\label{peva}
\end{align}
where
\begin{align}
&E^k=E^{k-1}+P_{EVA}^{k}\Delta k\\
&f_1 = \frac1K \sum_{k=1}^{K}P_{load}^k\label{f1}\\
&P_{load}^k = -P_{pv}^k-P_{wt}^k+P_{EVA}^{k}\label{pload}\\
&P^k=P_{base}^k-P_{pv}^k-P_{wt}^k+P_{EVA}^{k}\label{powerload}\\
&\sigma^{2}=\frac{1}{K} \sum_{k=1}^{K}\left(P^k-{P}_{ave}\right)^{2}\label{variance}\\
&{P}_{ave}=\frac{1}{K} \sum_{k=1}^{K}P^k\label{ap}
\end{align}
The objective function (\ref{pwgoal}) minimizes the overall cost of the DSO $F_1$, EVA $F_2$ and EV $F_3$, which includes the following components: 1) the variance in the load of the power grid ${\sigma^{2}}$ and the mean net load $f_1$; 2) the charging cost; 3) the SOH of EVs. And the solution variable is ${P_{EVA}^{k}}$.

$\alpha$ is the power grid fluctuation cost coefficient, which is $\$0.01/(kW)^2$. $\psi$ is the reward coefficient of $f_1$, which is positive and ensures the absorption of maximum renewable energy generation.  The positive constraint coefficient of the EVA cost $\upsilon$ ensures a minimal EVA cost. 

Based on the aggregate feasible second-order approximation region \cite{wen2022aggregate}, the EVA energy constraint is $\forall k_1 \in[1, K], \forall k_2 \in$ $\left[0, k_1-1\right]$ shown in (\ref{E}).
$\underline{e_n^{k_1/k_2}}$ is the $n$-th EV minimum energy at timeslot $k_1/k_2$. $\overline{e_n^{k_1/k_2}}$ is the $n$-th EV maximum energy at timeslot $k_1/k_2$. $\underline{p_n^k}$ and $\overline{p_n^k}$ respectively denote the maximum and minimum power of the $n$-th EV at timeslot $k$. $E^k$ is the EVA energy at timeslot $k$. $P_{EVA}^{k}$ is the EVA power at timeslot $k$. $\Delta k$ is the length of a one-time interval. The constraint can be intuitively interpreted as limiting the energy change between two time points. For example, the right part of the inequality means that the energy change of the EVA from $k_2$ to $k_1$ does not exceed the energy change from the lower energy limit of $k_2$ to the upper limit of $k_1$ or the consuming energy is kept from $k_2$ to $k_1$ with the maximum power. The left part can be explained similarly.

The tie-line power constraint is described in (\ref{ptie}). $\underline{P_{grid}}$ is the minimum tie-line power.
The transformer capacity constraint is described in (\ref{ptie}), where $S$ is the transformer capacity, which is 4000kVA. $cos\phi$ is the power factor, which is 0.8.
The power of EVA constraint is shown in (\ref{peva}) and is subject to the restrictions stipulated by (\ref{E}).
$\overline{P_{EVA}^k}$ and $\underline{P_{EVA}^k}$ are the maximum discharging power and the maximum charging power of the EVA, respectively. $\overline{E^k}$ and $\underline{E}^k$ are the maximum and minimum EVA energy in the timeslot $k$ respectively.
In (\ref{pload}), $P_{load}^k$ is the net load of the microgrid at the timeslot $k$, which is negative. Because renewable energy generation is negative, the smaller the $P_{load}^k$, the more renewable energy is consumed.
$P^{k}$ is the power load at the timeslot $k$ calculated as (\ref{powerload}).
The load variance $\sigma^{2}$ indicates the stability of the power grid load. The smaller its value is, the flatter the grid load is. The mathematical model of the load variance is shown as (\ref{variance}), $K$ is the number of timeslots.
$P_{ave}$ is the average of the total power load during the day, which is shown as (\ref{ap}).
$c^{k}$ is the time-of-use (TOU) tariff at the time slot $k$.
The degradation cost of an EV battery is $c_{bat}=\$300/kWh$, and the labor cost for battery replacement is $c_{l}=\$240$. The end-of-life SOH is set to be $80\%$. The EV battery is considered scrapped when the SOH drops below $80\%$. 
  
\section{MACPO and PoS-based Multi-Stakeholder Hierarchical V2G Coordination Strategy }
\subsection{MACPO-based EVA Coordination}
In MACPO-based EVA coordination, the agent interacts with an environment through a sequence of observations, actions, and rewards. The agent's goal is to select actions in a manner that maximizes the cumulative future reward\cite{mnih2015human}.
Taking into account the constraint of SOC, we formulate the EV coordination problem as a tuple of a constrained Markov decision process $\left(\mathcal{N,S},\mathcal{A},P,\mathcal{R},\mathcal{C},d\right )$.
A detailed explanation of the MACPO formulation for the large-scale V2G continuous charging and storage coordination problem is provided, and the critical elements of the tuple are described below.

\paragraph{Agent}  EVA constitutes a set of agents $\mathcal{N}=\left \{ 1,2,...N \right \} $, which gradually learns how to improve its retail charging distribution decisions using experiences from its repeated interactions with the environment.

\paragraph{Environment \& State}
The environment consists of EVs, power grid, and RES (wind power and PV power), with all of which the EVA interacts. $\mathcal{S}$ is a state space, and the agent state at timeslot k is  $s_k \in \mathcal{S}$ 
\begin{align}
    s_k=\{P^{k-23}, ..., P^k, E^k, {\sigma^{2}_k}, c^k, {\lvert E^k-E^{k-1} \rvert} \}\notag
\end{align}
encapsulates five types of information:

(1) $P^{k-23}, ..., P^k$ denotes the load values of the power grid in the past 24 hours before the time slot $k$;

(2) $E^k$ represents the EVA energy in the time slot $k$;

(3) ${\sigma^{2}_k}$ indicates the load variance in the time slot $k$.

(4) $c^k$ indicates the TOU tariffs in the time slot $k$.

(5) ${\lvert E^k-E^{k-1} \rvert}$ indicates the change in EVA energy at the time interval $k$.

In the environment, there is a transition dynamics distribution $P:\mathcal{S}\times\mathcal{A}\to \left [ 0,1 \right ] $ with conditional transition probability $p\left(s_{k+1} \mid s_{k}, a_{k}\right)$, satisfying the Markov property, i.e. $p\left(s_{k+1} \mid s_{k}, a_{k}\right)=p\left(s_{k+1} \mid s_{1}, a_{1}, \ldots, s_{k}, a_{k}\right);$ 

\paragraph{Action}
Given the state $\mathcal{S}$, agent $i$ will select an action $a_k^i$ at the timeslot $k$ from the joint action space $\mathcal{A}=\prod_{i=1}^{N}\mathcal{A}^i$.
\begin{align}\label{action}
    &\quad a^i_{k}=P_{EVA}^{k}
\end{align}

The action $a_{k}^i$ represents the charging/discharging power of the EVA at the timeslot $k$. Let $a_k^i$ be positive when EVA charges and negative when discharging. In addition, we assume that the V2G equipment provides continuous charging/distribution power.

\paragraph{Reward}
As encouragement from the environment in the reward space $\mathcal{R}$: $\mathcal{S} \times \mathcal{A} \rightarrow \mathcal{R}$, the reward is the goal of optimization for DSO. To avoid sparse rewards during the agent learning process, the following functions highlight the rewards and penalties for every action undertaken by the agent when scheduling EVA charging/discharging.
\begin{align}
r^i=\frac{1}{F_1+ F_2+ F_3}\label{rr}
\end{align}
where $r^i \in \mathcal{R}$ is the reward of the agent $i$, and the goal is to maximize (\ref{rr}).

\paragraph{Cost}
The environment returns the cost $c^i\in\left [ 0,1 \right ] $ for $i-th$ agent in the cost space ${\mathcal{C}}$: $\mathcal{S} \times \mathcal{A} \rightarrow \mathcal{C}$ as a signal of whether a selected action will overcharge the electrical battery and cause potential hazards.
If any of the constraints (\ref{E})-(\ref{peva}) are not satisfied, then the cost $c^i=1$; when all constraints are satisfied, $c^i=0$.

In time slot $k$, the agent $i$ observes the local environment state $s_k^i$ and selects a specific action $a_k^i$, drawn from its policy $\pi^i\left ( \cdot |s_k^i \right ) $ . After performing the action, the environment returns a collective reward $r_k$ and a respective cost $C_k^i$ while moving to a new state $s_{k+1}$. Denoting the reward as the degree of mitigating load fluctuation, as well as the economic benefits for EV users, and the cost as an indicator of whether the behavior tends to render SOC cross-border.

To solve the constrained Markov decision process, we combine the trust region policy optimization with constrained policy optimization, then generalize to Multi-agent policies, similar to the work of \cite{gu2021multi}. Apart from the main idea, we incorporate a safety-checking process into the policy to prohibit any SOC violation.

Denoting the joint policy as $ \pi(\cdot |s_k)=\prod_{i=1}^{N}\pi_i(\cdot^i|s_k) $, the joint action as $\boldsymbol{a}_k=\left ( a_k^1,a_k^2,...a_k^N \right )$. Then, the action-state value function, state value function, and advantage function in terms of reward are defined to be the following equations: 
\begin{align}
    &Q_{\pi}(s,\boldsymbol{a})\doteq \mathbb{E}_{\tau\sim\pi}\left [\sum_{k=0}^{k=K} \gamma ^kR_k(s_k,\boldsymbol{a_k}) |s_0=s,\boldsymbol{a_0}=\boldsymbol{a}\right ] \notag\\ 
    &V_{\pi}(s)\doteq \mathbb{E}_{\boldsymbol{a}\sim\pi}\left [ \sum_{k=0}^{k=K} \gamma ^kR_k(s_k,\boldsymbol{a_k}) |s_0=s\right ] \notag\\
    &A_{\pi}(s,\boldsymbol{a})\doteq Q_{\pi}(s,\boldsymbol{a})-V_{\pi}(s) 
\end{align}
where $\gamma$ denotes the discounted factor and $\tau$ is the trajectory from start to end. 
Changing the reward in the above definitions with the individual cost, we can obtain the cost counterpart for each agent, respectively, represented by $Q_{C,\pi}^i(s,a^i),\ V_{C,\pi}^i(s),\ A_{C,\pi}^i(s,a^i)$.

In multi-agent CMDP, aggregators are aimed at finding the optimal joint policy that maximizes the expected total reward 
$J(\pi)\doteq
\mathbb{E}_{\boldsymbol{\tau }\sim \pi}\left[\sum_{k=0}^{K} \gamma^{k} R_k\left(s_k, \boldsymbol{a}_{k}\right)\right]$ 
and keeping the expected total cost
$J_C^i(\pi)\doteq
\mathbb{E}_{\tau\sim\pi}\left[\sum_{k=0}^{k} \gamma^{k} C_k^i\left(s_k, \boldsymbol{a_k}\right)\right]$
within the bound $d^i$ for any agent $i$.
According to Kuba et al.'s work, the joint advantage function could be decomposed to the summation of subsets advantages\cite{kuba2022trust}.
For any state $s$, subset of agents $i_{1:m}\subseteq \mathcal{N} $, and joint action $\boldsymbol{a^{i_{1:m}}}$, the following identity holds
\begin{align}\label{17}
    A_\pi^{i_{1:m}}\left (s,\boldsymbol{a}^{i_{1:m}} \right ) =\sum_{j=1}^{m}A_\pi^{i_j}\left(s,\boldsymbol{a}^{i_{1:j-1}},a^{i_j}\right) 
\end{align}
where $A_\pi^{i_j}$ denotes the advantage of implementing $a^{i_j}$ over not doing it, with $\boldsymbol{a^{i_{1:j-1}}}$ having already been done  by the subsets of agents $i_{1:j-1}$.

Serving as a transition from single agent to multi-agent problem, Equation (\ref{17}) implies the positive assurance of $A_\pi$ as long as actions are taken sequentially in an arbitrary order and ensures $A_\pi^{i_j}\ge 0 $. Specifically, we can build the lower bound of the objective function 
\begin{align} \label{18}
    J\left (\widetilde{\pi } \right )\ge J\left (\pi \right ) + \sum_{m=1}^{N}\left[ \mathbb{E}_{\boldsymbol{a}^{i_{1:m-1}}\sim\boldsymbol{\widetilde{\pi}}^{i_{1:m-1}}}\left [ A_\pi^{i_m}\left ( s,\boldsymbol{a}^{i_{1:m-1}}, a^{i_m}\right ) \right ]\notag\right.
    \\
    \phantom{=\;\;}\left.-\nu D_{KL}^{max}\left ( \widetilde{\pi} ^{i_m},\pi^{i_m} \right) \right ] \qquad\qquad\qquad\qquad 
\end{align}
where$\quad\nu =\frac{4 \gamma \max _{s, a}\left|A_{C,\pi}\left(s, \boldsymbol{a}\right)\right|}{(1-\gamma)^{2}}$, and $\boldsymbol{\widetilde{\pi}}$ is the updated policy. We denote the max-KL divergence between the updated policy of agent $i_m$ and the old one as $D_{KL}\left ( \widetilde{\pi} ^{i_m},\pi^{i_m} \right)$.
The inequality indicates that if the agents' policies are optimized sequentially, we can guarantee a monotonic improvement of the joint performance $J\left (\widetilde{\pi } \right )$.
Then, to incorporate the cost constraints into the above problem formulations, a tight upper bound of $J_C^i\left(\widetilde{\pi}\right)$ was theoretically justified, extended from the reward objective (\ref{18}).
\begin{align}\label{lemma1}
    J_C^{i}(\widetilde{\pi}) &\leq J_C^{i}(\pi)+\mathbb{E}_{a^i\sim \widetilde{\pi^i}}\left[ A^i_{C,\pi}\left ( s,a^i \right ) \right]+\nu^{i} \sum_{i=1}^{N}  D_{KL}^{\max }\left(\pi^{i}, \widetilde{\pi}^{i}\right) \notag\\
    & \qquad \text{where}\quad\nu^{i}=\frac{4 \gamma \max _{s, a^{i}}\left|A_{C,\pi}^{i}\left(s, a^{i}\right)\right|}{(1-\gamma)^{2}} 
\end{align}

Equation (\ref{lemma1}) implies that the change of expected cost for every agent can be controlled by the surrogate cost $ L_C^i(\pi^i)$, if the KL-divergence between two policies is small enough, which provides a guide for updating joint policies, with a guarantee of satisfaction of monotonic safety constraints, in addition to the improvement of reward performance, that is, $J\left(\widetilde{\pi}\right)\ge J\left(\pi\right)$ and $J^i\left(\pi\right)\le c^i,\forall i\in \mathcal{N} $. Specifically, let agents solve the following optimization problem sequentially:
\begin{align}
    \widetilde{\pi^{i_m}} &=arg\max_{\pi} \mathbb{E}_{\boldsymbol{a}^{i_{1:m-1}}\sim\boldsymbol{\widetilde{\pi}}^{i_{1:m-1}}}
    \left [A_\pi^{i_m}\left(s,\boldsymbol{a}^{i_{1:m-1}},a^{i_m}\right) \right]\notag\\
    s.t.&\ J_C^{i_m}(\pi)+\mathbb{E}_{\boldsymbol{a}^{i_{1:m-1}}\sim\boldsymbol{\widetilde{\pi}}^{i_{1:m-1}}}\left [A_{C,\pi}^{i_m}(s,a^{i_m})\right]\le c^{i_h}, \notag\\        
    &\qquad\qquad\overline{D_{KL}} (\pi^{i_m},\widetilde{\pi^{i_m}})\le\delta
\end{align}
where $\overline{D_{KL}}$ is the expected KL-constraint, which is bounded by threshold $\delta$.

Note that we have already adopted the trust-region method, which relaxes the $D_{KL}^{max}$ with $\overline{D_{KL}}$, rather than penalizing both the objective and the cost-constrained equation. Since addressing $D_{KL}^{\max }$ as a penalty term could result in a dilemma, {\it{i.e.}}, if the coefficient of the penalty term is set too small, the agent might overlook the constraints. In contrast, the agent will stay still to satisfy the constraints, rendering slow performance improvement. Furthermore, the computational cost of the max KL-divergence must be considered. 
The detailed explanation of the derivation of MACPO can be found in Appendix~\ref{appb}.
Following the principles of Taylor expansion, the optimization problem can be approximated linearly or quadratically. To tackle this problem, we can initially focus on its Lagrange dual problem, which is convex in nature, enabling us to derive the primal solution swiftly~\cite{kuba2022trust}.

Training the agent neural network in MACPO can ensure that the policy is within the SOC boundary, with a higher probability of not violating the constraint. However, the agent cannot guarantee the safety of the entire participant during training, even in execution, which is not allowed for EV batteries, since overcharging will likely cause significant degradation in performance or combustion. Based on these considerations, we introduce a safety check mechanism that prohibits cross-border behavior. If one action causes SOC beyond the constraint, the action would be clipped to the boundary of its feasible region immediately. Moreover, although the unsafe step would not be implemented, the cost still exists as a guide, warning the policy and the value network of possible dangerous violations.
\begin{figure*}[htbp]
\centering
\includegraphics[width=0.9\textwidth]{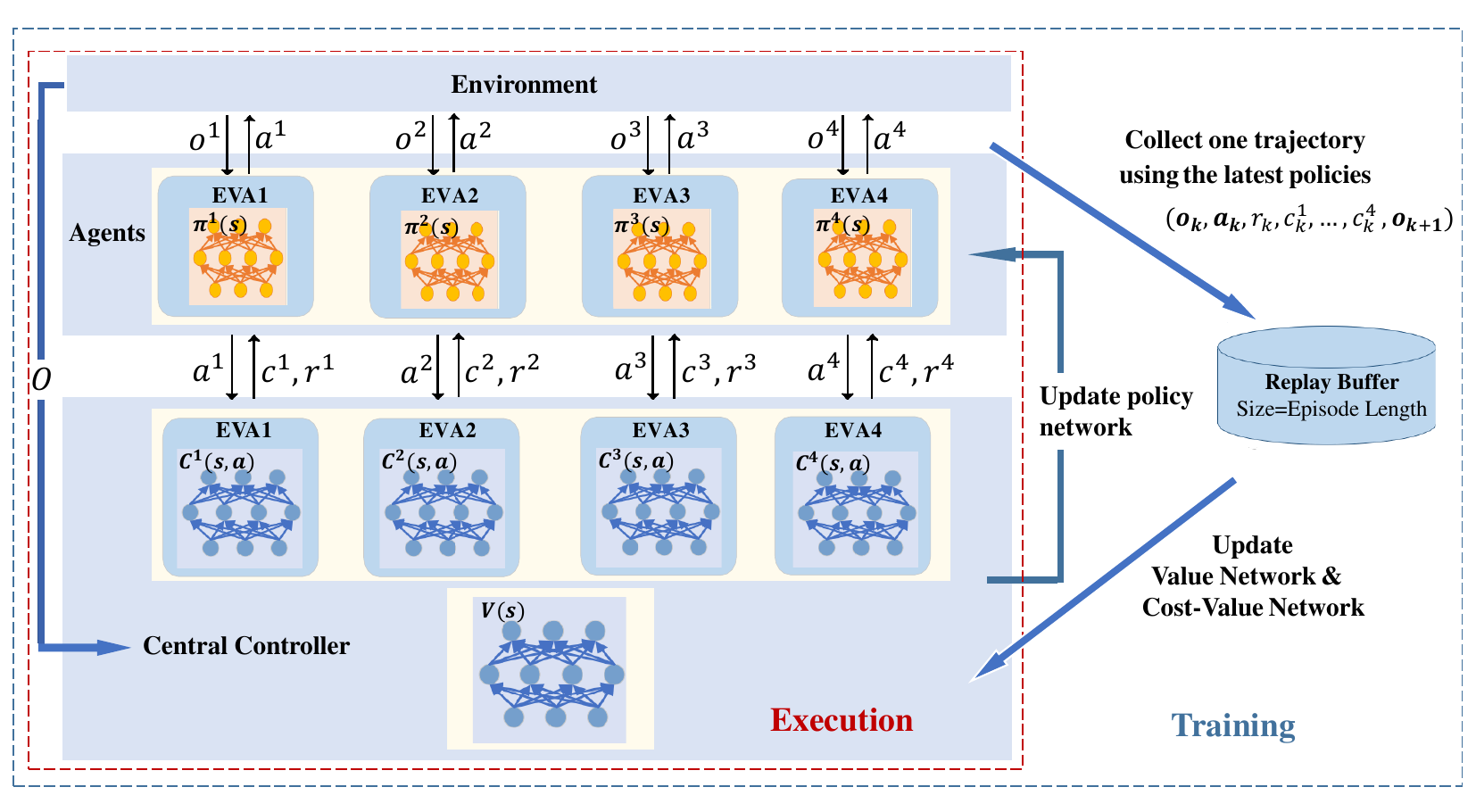}
\caption{The MACPO architecture for training and execution.}
\label{fig:macpoArch}
\end{figure*}

The architecture of hierarchical MACPO-based EVA coordination is illustrated in Fig.~\ref{fig:macpoArch}, which includes three types of neural networks:
A global value network $V\left (s,\boldsymbol{a};w \right ) $ to estimate the advantage of the selected joint action in a global state, a cost value network $C\left ( s, a  \right ) $ for each agent to evaluate violation of constraints, which possesses the same structure as above, and an individual policy network $\pi\left ( s\right ) $ mapping the local state into a probability distribution over the action space.
All networks are implemented using fully connected layers with the same input size, and the rectified linearity units (ReLU) for the hidden layer are employed. 
The cost value and policy network have one output neuron, whereas the global value network generates a vector of length N, with each neuron corresponding to an agent. 
The output layer in the policy network is a sigmoid layer with bounded continuous actions within $\left [ -1,1 \right ] $. 

During one training episode, which begins with the sampling stages of the experiences, the agents interact with the environment for $20$ epochs, corresponding to 20 hours in the context of our problem model. Specifically, agents observe the state $s^i_k$ and feed it into the policy network $\pi\left (s^i_k\right) $, generating action. The environment then provides rewards and costs for the action and generates a new state. If those steps are iterated more than 20 times, a complete trajectory can be collected and saved in the replay buffer. First, let the value and cost-value network compute the estimated advantages depending on the sequential transitions in the stochastic optimization stage. Using these data, the policy network can be updated with a conjugated gradient and backtracking line search, and the two value-based networks can be updated through a temporal difference algorithm.

\subsection{PoS-inspired EVA Power Allocation Algorithm}
With the increasing demand for energy efficient, secure, and decentralized solutions to manage transactions in smart grids, the PoS algorithm has attracted attention due to its energy efficiency and enhanced security features.
\paragraph{EV User Selection} EV users signal their intention to participate by locking a portion of their battery energy as collateral. The selection process considers factors such as the amount of battery energy, the age of the batteries, and a randomization factor to ensure fair and unbiased participation.
\paragraph{Power Proposal} The selected EV user proposes a new charging/discharging power containing validated transactions. To obtain the individual EV power, the PoS assigns an energy weight $\zeta$ to the battery energy of each EV user in proportion to the total battery energy of the EV user. Thus, the general principle remains constant: EV users with more battery energy are more likely to charge and discharge.
\paragraph{Validation and Consensus} Other EV users in the network verify the authenticity of the proposed charging/discharging by checking if the proposer has met the selection criteria and constraints. A consensus mechanism, such as Byzantine Fault Tolerance, ensures that EV users agree on the validity of the proposed charging/discharging before adding it to the strategy.
\paragraph{Rewards and Penalties} EV users receive rewards in the form of cost savings in charging and mitigating battery aging for their contributions to the power grid. PoS encompasses punitive measures against malicious behavior, such as deviating from the power plan or early EV usage, resulting in partial battery energy loss. The deterrent system discourages validators from dishonestly using a deterrent system and preserves the security and integrity of the network.
The energy weight $\zeta$ can be defined as:
\begin{align}
&\zeta_n=\left\{\begin{array}{cc}
\frac{(e_{n}^{k}-e_{n}^{k-1})}{\overline{e_{n}^{k}}-e_{n}^{k-1}} &P_{EVA}^{k}>0 \\
\frac{(e_{n}^{k-1}-e_{n}^{k})}{e_{n}^{k-1}-\underline{e_{n}^{k}}} &P_{EVA}^{k}<0
\end{array}\right.\label{kappa}\\
&e_{n}^{k}=e_{n}^{k-1}+{\eta p_{n}^{w,k^{*}}}\Delta k\label{evsoc}
\end{align}
subject to
\begin{align}
&\overline{P_{n,dis}^{w,k}}\leq p_{n}^{w,k^{*}}\leq \underline{P_{n,ch}^{w,k}}\\
&\underline{SOC_{n}^{k}}{Q^w_n}\leq e_{n}^{k}\leq \overline{SOC_{n}^{k}}{Q^w_n}\label{deps}\\
&P_{EVA}^{k}=\sum\limits_{n \in \mathcal{N}}{{\eta p_{n}^{w,k^{*}}}\Delta k}/{Q^w_n}\label{powerall}
\end{align}
$p_{n}^{w,k^{*}}$ represents the charging/discharging power of the $n$-th battery in the timeslot $k$ during the $w$-th equivalent full cycle. $\eta$ is the charging/discharging efficiency of the $n$-th battery. $Q^w_n=Q_nSOH^i_n$ is the battery capacity of the $n$-th EV after the $w$-th equivalent full cycle. $SOH^w_n$ denotes the state of health of the $n$-th EV battery after the $w$-th equivalent full cycle. $\Delta k$ is the time interval, which is one hour.
$\underline{SOC_{n}^{k}}$ and $\overline{SOC_{n}^{k}}$ are the minimum and maximum SOCs for $n-th$ EV at the time slot $k$, respectively. In particular, when $\overline{e_{n}^{k}}=e_{n}^{k-1}$ and $P_{EVA}^{k}>0$, $p_{n}^{k^{*}}=0$; when $\underline{e_{n}^{k}}=e_{n}^{k-1}$ and $P_{EVA}^{k}<0$, $p_{n}^{k^{*}}=0$.

\begin{figure}[h]
\centering
\includegraphics[width=0.4\textwidth]{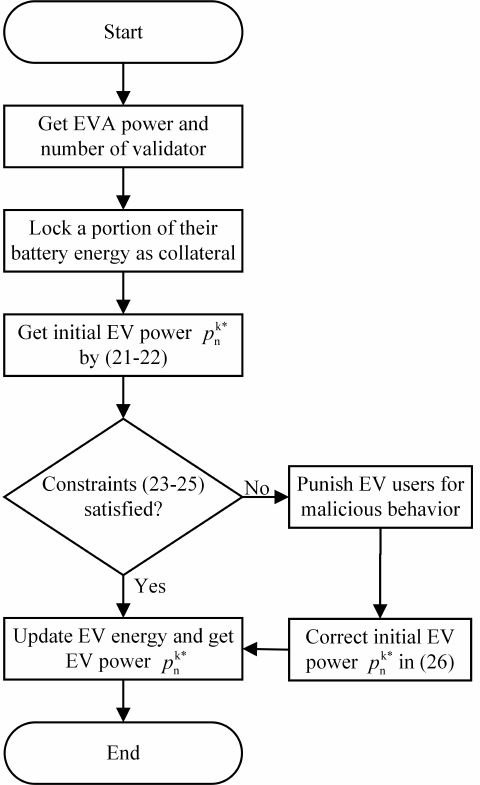}
\caption{Flowchart of the PoS-inspired EVA power allocation algorithm.}
\label{fig:flow}
\end{figure}

After calculating $p_{n}^{k^{*}}$, a safety check correction is needed to obtain the true charging/discharging power $p_{n}^{k}$. During scheduling, the upper and lower limits of the output power of each EV are used to correct the EV energy because the maximum charging/discharging power should not be exceeded. An individual electric vehicle should meet power constraints and the charging / storage power $p_{n}^{k}$ is corrected as follows.
\begin{align}\label{soc1}
&p_{n}^{k}= \begin{cases}p_{n}^{k^{*}} & \overline{p_{dis}} \leqslant p_{n}^{k^{*}} \leqslant \overline{p_{ch}} \\
\overline{p_{ch}} & P_{EVA}^{k}>0, p_{n}^{k^{*}}>\overline{p_{ch}} \\
\underline{p_{dis}} & P_{EVA}^{k}<0, p_{n}^{k^{*}}<\overline{p_{dis}}\end{cases}
\end{align}
where $\overline{p_{ch}}$ and $\overline{p_{dis}}$ are the maximum charging/discharging power of the EV, respectively. The allocation of EVA power is presented in Fig.~\ref{fig:flow}.
The energy of each EV can effectively converge, and all reach the desired energy before their departure time, solving the redistribution problem caused by the initial energy difference.
\section{Simulation Results}

\subsection{Test System and Implementation}
\begin{figure}[h]
\centering
\includegraphics[width=0.45\textwidth]{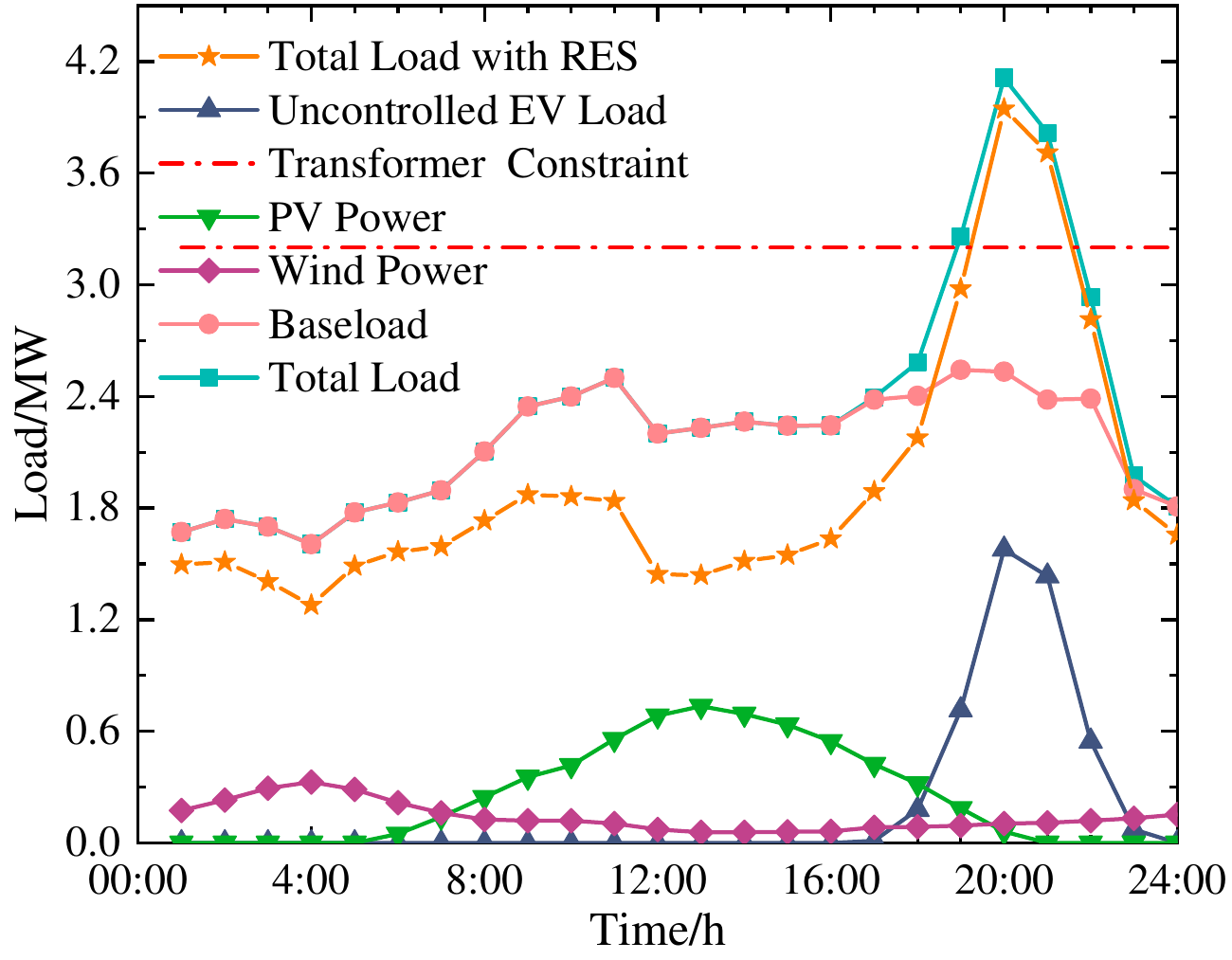}
\caption{Daily load profiles in a microgrid.}
\label{fig:baseload}
\end{figure}

The simulation results tackle the Electric Vehicle Aggregator (EVA) scheduling problem over a single day with hourly resolution. The EVA manages the charging of 509 EVs to mitigate charging load peaks. Fig.~\ref{fig:baseload} displays the total load, the baseload, and the uncontrolled EV load for the day in question. The total load profile combines the uncontrolled EV load with the baseload profile. The total load exceeds the transformer capacity with and without renewable energy sources (RES). When RES is integrated, the total load of the microgrid is balanced by contributions from EVs, baseload, photovoltaic systems (PVs) and wind power.

The control parameters of the EV are $p_{ch}^{max}= 6 kW$, $p_{dis}^{max}= -6 kW$, $Q_n=24 kWh$ for all EV. The detailed EV load model is shown in \cite{zhang2021optimal}. The arrival time is sampled from $N(18, 1^2)$ and is bounded between 15 and 21. Its distribution $N(8, 1^2)$ for the departure time is bounded between 6 and 10. The SOC of $n-th$ EV $SOC_n$ bounded between 0.2 and 0.8 is sampled from $N(0.5, 0.1^2)$. It should be noted that in the mentioned test systems, the EVA is scheduled from timeslots 15:00 to 10:00. To improve the users' participation rate, the enrollment incentive ($\$560$ per customer) is set for users' discharging compensation to compensate users for discharging~\cite{9384297}. It should be noted that the hierarchical coordination strategy does not rely on any knowledge of the distributions of these random variables. Thus, the proposed coordination can be transferred to different modeling mechanisms.

\begin{figure}[h]
    \centering
\includegraphics[width=0.5\textwidth]{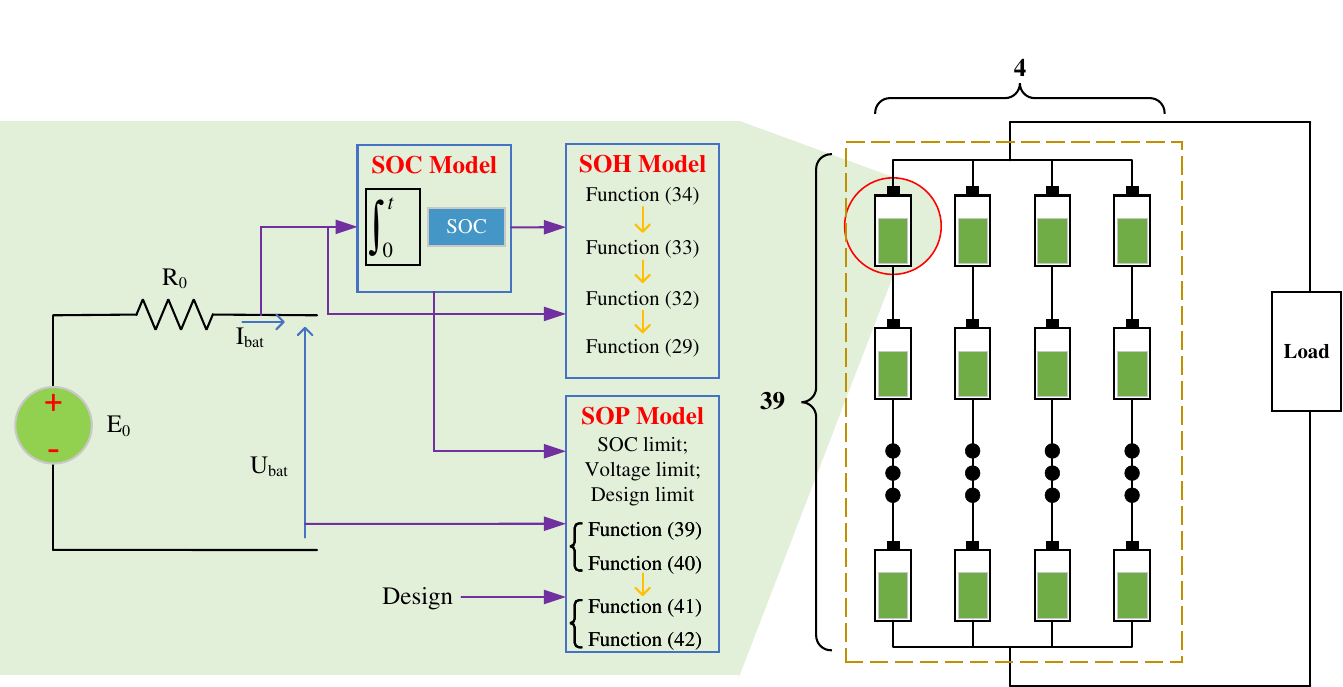}
    \caption{Electric vehicle battery simulation system.}
    \label{fig:simu}
\end{figure}

The topology of the series-parallel battery array is shown in the Fig.~\ref{fig:simu}. The model connects battery cells in series and in parallel to form a battery module. There are 39 battery cells connected in series and 4 parallel branches. The rated voltage of the battery cell is 3.3V, the rated capacity is 2.3Ah, and the initial equivalent full-cycle number of the battery is 50. Depending on the battery parameters, it is modeled and simulated in the MATLAB Simulink, as shown in Fig.~\ref{fig:simu}. The detailed simulation environment with battery conditioning is shown in Appendix~\ref{sebc}.

\subsection{Training Process of MACPO-based EVA Coodination}
\begin{figure}[h]
\centering
\includegraphics[width=0.45\textwidth]{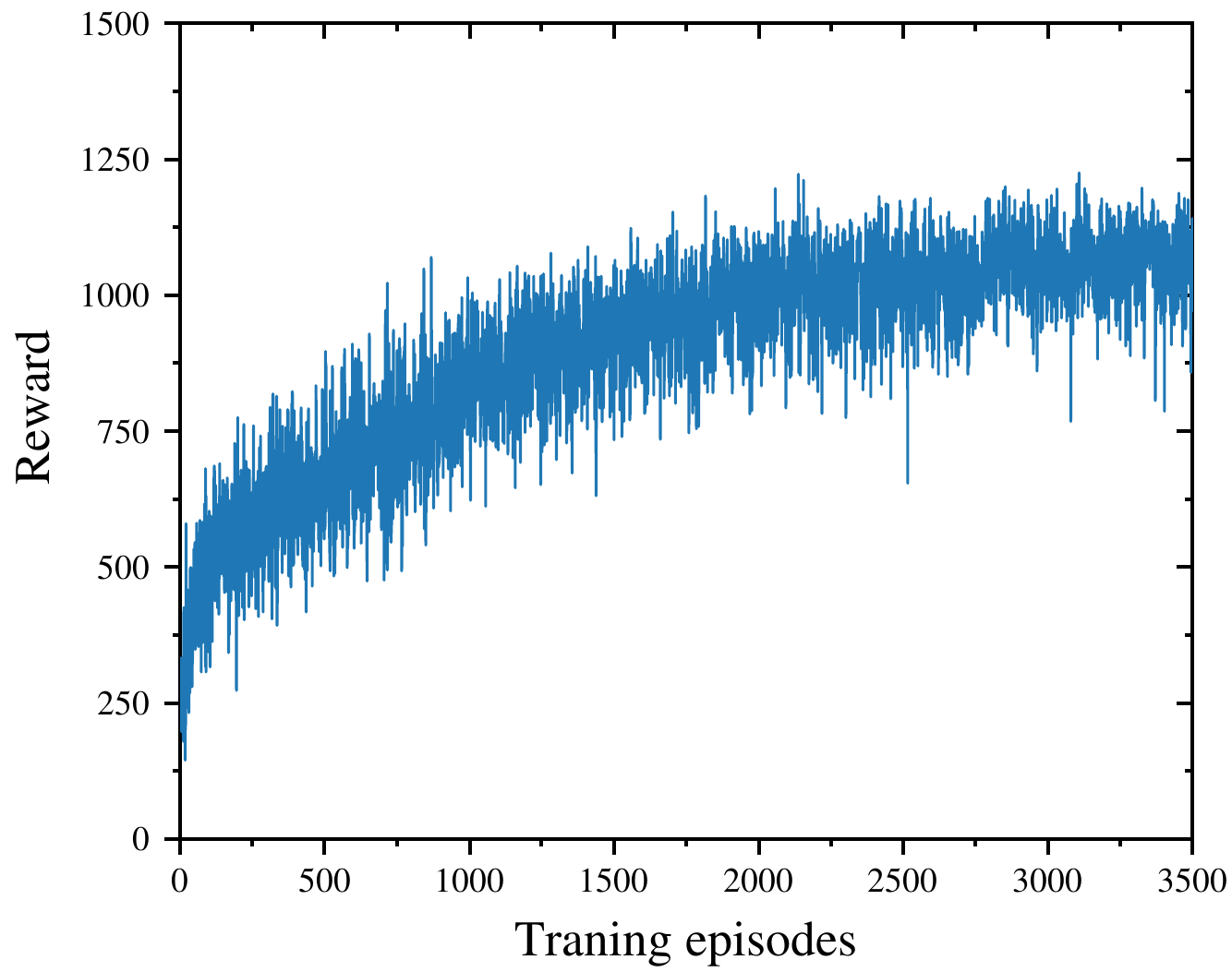}
\caption{The evolution of the rewards during the training process.}
\label{fig:train}
\end{figure}

In the training, the upper limit of the constraint was set at 0.1. 
Adam optimizer is used to optimize the value and cost value network, with the learning rate selected as $lr=5e^{-4}$. In contrast, the policy network is updated through conjugate gradient and backtracking line search for 10 iterations, with the step size being 0.1.
We trained the model for 3500 episodes, with 20 hours per episode, using 5 parallel environments to accelerate the sampling process. 
Since we use generalized advantage estimation (GAE), which requires a complete trajectory to compute the advantage, the replay buffer size $B$ equals the length of episode 20. The smoothing parameter $\lambda$ and the discounted factor $\gamma$ used in GAE are selected as 0.95 and 0.99, respectively.
During the training process, the reward and load variance in episodes 3500 are calculated and illustrated in Fig.~\ref{fig:train}.
As demonstrated in Fig.~\ref{fig:train}, the charging/discharging schedule is randomly selected during the initial learning stages as the EVA gathers more experiences by randomly exploring different, not necessarily profitable, actions. However, as the learning process progresses and more experiences are collected, the reward keeps increasing, eventually converging around 1141.2 with small oscillations. The result demonstrates that the proposed strategy successfully learns a policy to maximize the reward.

\subsection{MHVCS Performance Benchmark}

To evaluate the performance of the proposed MHVCS, the following five baseline strategies with the uncontrolled charging control, optimal charging control, charging/discharging control with minimization of grid load fluctuation,  with minimization of cost, and with a single PPO agent:
\begin{itemize}
\item Baseline 1 (BL1), Uncontrolled charging: The user does not consider the price of electricity and the load on the power grid. When they return home, they charge the battery at maximum power until it is full.
\item Baseline 2 (BL2), Optimal charging: The user considers the SOC of the battery and the load on the power grid. They charge according to the EVA's instructions until the battery is full.
\item Baseline 3 (BL3), charging/discharging with minimization of grid load fluctuation: The user considers the level of grid load and charges/discharge by minmizing the fluctuation of the grid load without considering battery aging until the battery is full.
\item Baseline 4 (BL4), charging/discharging with minimization of cost: The user considers TOU tariffs and charges/discharges to minimize the cost without considering the battery degradation.
\item Baseline 5 (BL5), Charging and discharge by hierarchical V2G coordination with the single PPO agent considering the SOX of the battery. The goal is equation (\ref{rr}). 
\end{itemize}
Three evluation indices (EI), the one-year SOH, the one-day load variance (LV), and the one-day EV total cost, are used to benchmark the performance of MHVCS for the multi-stakeholders
for the five baseline strategies, are shown in Table~\ref{tab:soh}. 

\begin{figure}[h]
\centering
\includegraphics[width=0.45\textwidth]{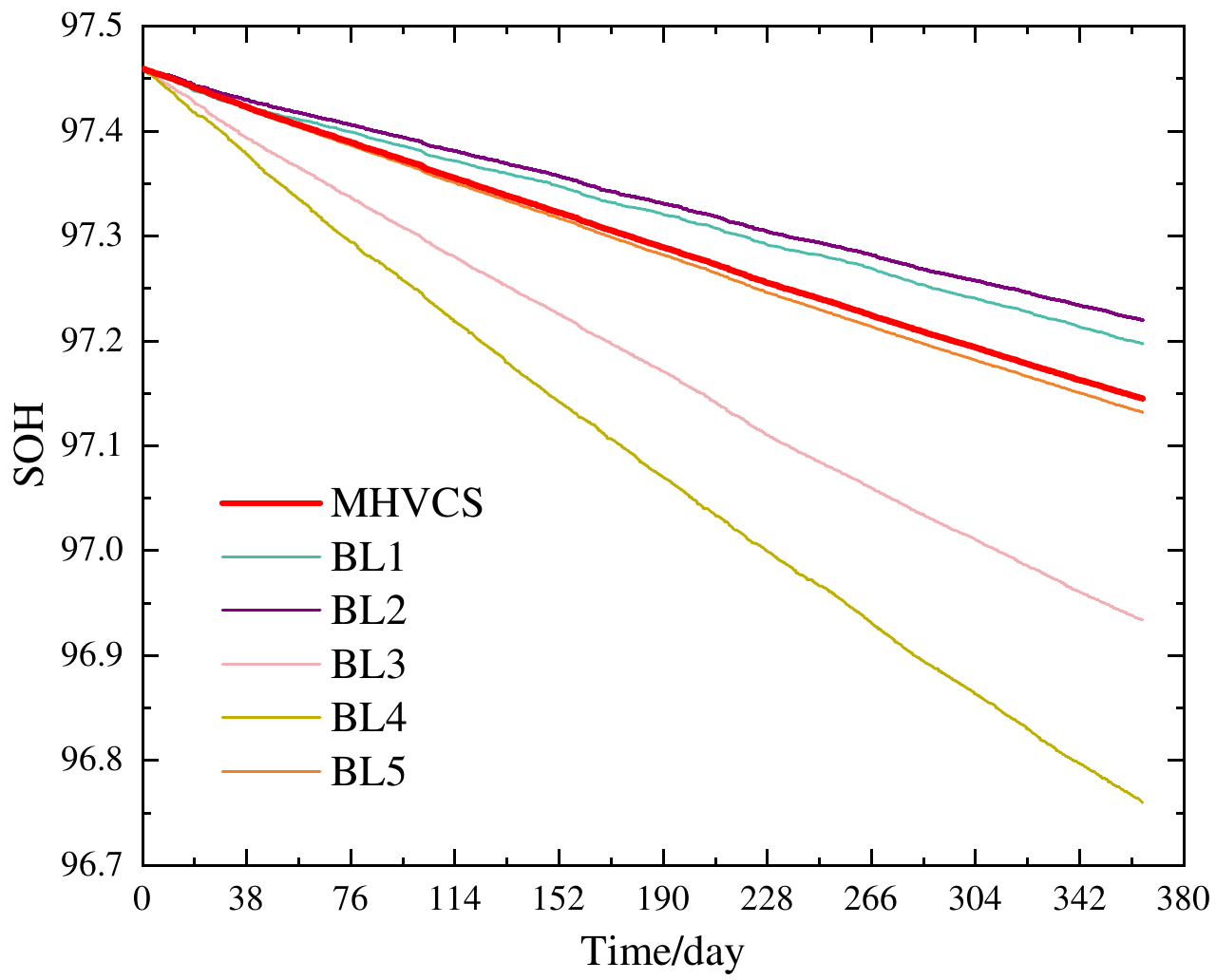}
\caption{The SOH curve under MHVCS and five baseline strategies after one year.}
\label{fig:soh}
\end{figure}

\paragraph{One-year SOH} The SOH, after one year of operation, is used to measure the performance of battery degradation under the coordination strategy. The SOH curves under MHVCS and five baseline strategies after one year of operation are shown in Fig.~\ref{fig:soh}. The initial value of SOH was 97.46. The battery degradation in MHVCS is not much different from the uncontrolled and optimal charging strategy in the BL1 and BL2 strategies and is better than the charging/discharging control in the BL3, BL4 and BL5strategies, as shown in Fig.~\ref{fig:soh}.

\begin{figure}
\centering
\includegraphics[width=0.45\textwidth]{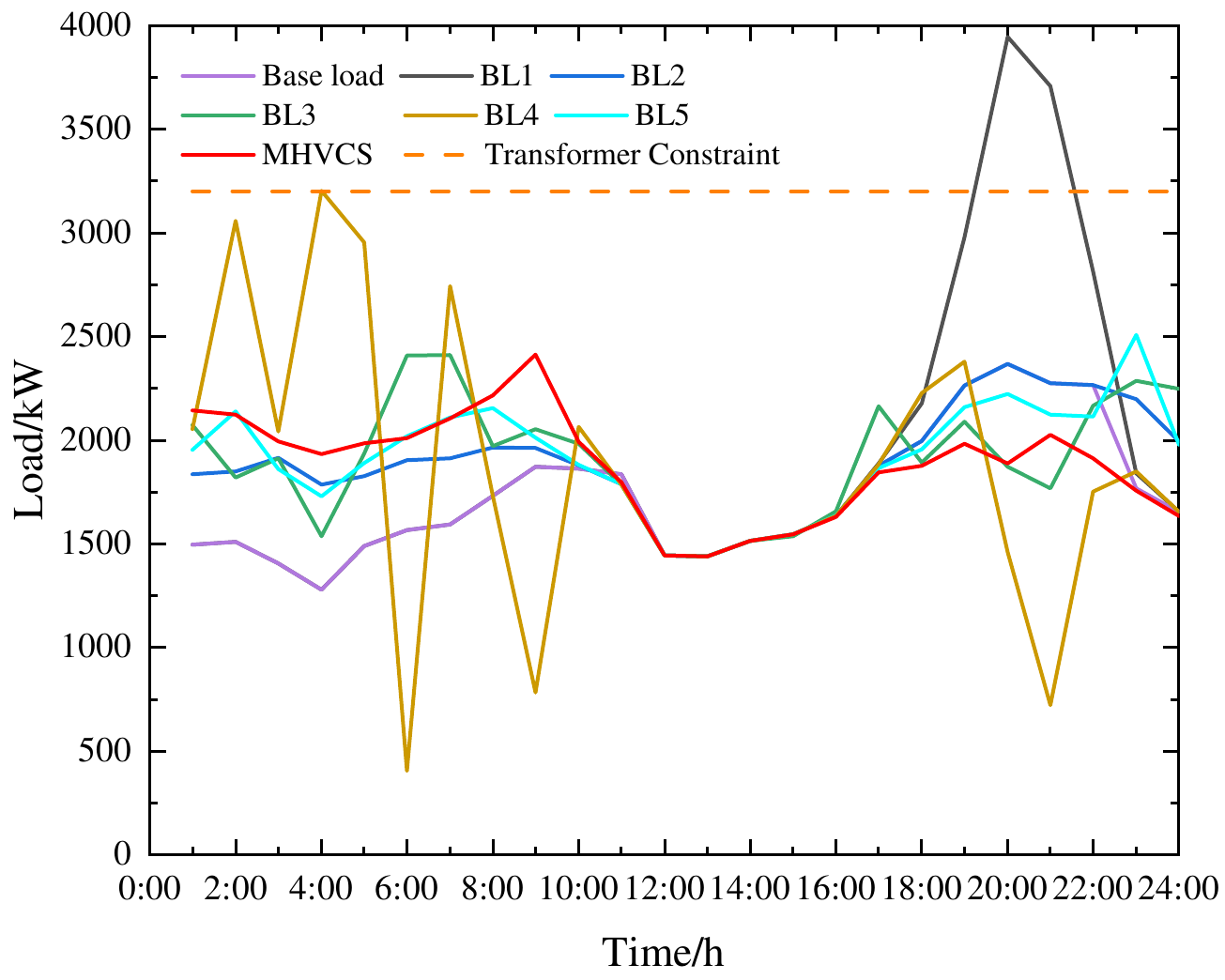}
\caption{Load profile under MACPO-based MHVCS and five baseline strategies in a day.}
\label{fig:load}
\end{figure}

\paragraph{One-day load variance} The performance of peak shaving and valley filling is evaluated with load variance in scheduling coordination, as shown in Fig.~\ref{fig:load}.
The load variance at each timeslot is calculated using the load profile of the previous 24 hours using equation (\ref{variance}). 
In Fig.~\ref{fig:load}, the optimal load profile generated by MHVCS is much flatter than the BL1 strategy with uncontrolled charging and the BL2 strategy with the optimal charging strategy since MACPO can leverage the charging/discharging control in the four EVAs to mitigate the peak load. In fact, both MACPO-based MHVCS and BL5 strategy demonstrate the capability to even out the load pattern by adhering to the EVA schedule based on a deep reinforcement learning strategy. This involves distributing the EVA's charging demand to the low points in the load profile. However, MHVCS with MACPO multi-agent control is more efficient at peak shaving and has less load variation in adjacent hours than the single-agent control in the BL5 strategy. It is also noted that the BL3 strategy, which aims at minimizing load fluctuation, also shows an excellent tendency to flatten the load curve, while still inefficient in suppressing load variation on a shorter time scale.
At the arrival time of timelot 15, the coordination schedule starts to connect the EVA to the power grid. At the departure time of timeslot 10, the departure load variance becomes 71312.8${kW}^2$. During the scheduling period, the MACPO agents provide charging/discharging actions according to the states $\mathcal{S}$. Specifically, EVA discharges at the load peak timeslot 16:00-21:00 and is charged at the remaining time in the scheduling period. 

\paragraph{One-day EV cost} The EV cost $C$ and the battery degradation cost $C_{b}$ \cite{singh2020cost} are shown as:
\begin{align}
&C=C_{ch}+C_{b},\\
&C_{b}=\sum_{n=1}^{N}{(C_{bat}+\frac{C_{l}}{1-SOH_{min}}){(1-SOH^w_n)Q_n}},\label{cbattery}
\end{align}
where $C_{ch}$ is the charging cost and $C_{l}$ is the labor cost.
The degradation cost of an EV battery is $C_{b}$=300 kWh, and the labor cost for battery replacement is $C_{l}$=240. The EV battery is considered scrapped when the SOH drops below $80\%$,{\it{i.e.}} the end-of-life (EOL) SOH is 80\%.

\begin{table}[h]
 \centering
 \caption{\textsc{EIs under MHVCS and five baseline cases.}}
 \label{tab:soh}
 \setlength{\tabcolsep}{0.5pt}
 \begin{tabular}{ccccccc}
 \toprule
\makecell[c]{EIs} &\makecell[c]{MHVCS} &\makecell[c]{BL1} &\makecell[c]{BL2} &\makecell[c]{BL3} &\makecell[c]{BL4}&\makecell[c]{BL5}\\ 
\midrule
SOH$^a$ &97.15   & 97.20  &97.22   & 96.93 & 96.76 &97.13\\
LV$^b$ &62693.3  &508479.1   & 85010.7   & 67074.3 &485071.2 &71312.8\\ 
EV Cost$^c$ &1628.4  &5524.9   & 2271.4   & 2085.6 &-708.6 &2023.1\\
\bottomrule
 \end{tabular}
\begin{tablenotes}
   \item[*] $^a$ one-year SOH, $^b$ one-day load variance, $^c$ one-day EV total cost 
\end{tablenotes}
\end{table}
The proposed MHVCS can reduce the load variance, the battery degradation, and the total cost as shown in Tables~\ref{tab:soh} and ~\ref{tab:cost}. The one-year SOH from MHVCS is not much different from the uncontrolled and optimal charging BL1 and BL2 strategies. However, it is significantly larger than the BL3 and BL4 strategies. 
Regarding the grid load fluctuation, the LV of MHVCS is slightly higher than the BL3 charging/discharging strategy with minimum grid load fluctuation. However, it is significantly less than the LVs of the BL1, BL2 and BL4 strategies.
MHVCS has a lower EV cost than the BL1, BL2, BL3, and BL5 strategies but is comparatively more costly than the BL4 strategy. However, MHVCS exhibits significantly lower grid load variance than the BL4 strategy.

\subsection{DSO Cost} 
The DSO cost includes the charging, degradation, and load fluctuation costs for the EV and DSO respectively. The DSO cost decomposition is shown in Table~\ref{tab:cost}. The electricity selling companies should minimize the peak-to-valley difference and reduce grid fluctuations to maintain grid stability. Therefore, the electricity selling company will bear part of the cost of grid fluctuation, which can be expressed as $0.0001\sigma^{2}$\cite{}.
The DSO cost includes the charging, degradation, and load fluctuation costs for the EV and DSO respectively. The DSO cost decomposition is shown in Table~\ref{tab:cost}. The electricity selling companies should minimize the peak-to-valley difference and reduce grid fluctuations to maintain grid stability. Therefore, the electricity selling company will bear part of the cost of grid fluctuation, which can be expressed as $0.01\sigma^{2}$\cite{DWJS202402021}.
The grid load fluctuation cost factor is 0.01kWh$^2$.

Regarding DSO costs, the MHVCS charging cost is only slightly higher than the BL4 strategy, which aims to optimize the charging cost, while being lower than the BL1, BL2, and BL3 strategies. Compared to the BL5 strategy with the single agent algorithm PPO, the MACPO-based MHVCS allows for more precise charging/discharging control with less charging, degradation, and load flucuation costs.
\begin{table}[h]
 \centering
 \caption{\textsc{one-day DSO cost decomposition under MHVCS and five BLs strategies.}}
 \label{tab:cost}
  \setlength{\tabcolsep}{0.5pt}{
 \begin{tabular}{ccccccc}
 \toprule
\makecell[c]{Cost} &\makecell[c]{MHVCS} &\makecell[c]{BL1} &\makecell[c]{BL2} &\makecell[c]{BL3} &\makecell[c]{BL4}&\makecell[c]{BL5}\\ 
\midrule
Charging  &1294.6  &5196.9   & 1945.7   & 1726.0 &-1088.1 &1686.9\\
Degradation  &333.8  &328.0  & 325.7   & 359.6 &379.5 &336.2\\
Load Fluctuation  &626.9  &5084.8   & 850.1   & 670.7 &4850.7 &713.1\\
DSO  &2255.3  &10609.7   & 3121.5   & 2756.3 &4142.1 &2736.2\\
\bottomrule
 \end{tabular}}
\end{table}

\subsection{SOC and SOP Constraints}
SOC is the critical parameter for properly controlling the EV and securing power responses to changes under operating conditions. The SOC constraints will ensure that the hierarchical coordination strategy satisfies the driving demand of EV users.  SOC constraints can reduce the long-term capacity fade rate and achieve a higher number of equivalent full cycles or a higher cumulative discharge capacity over the battery's useful life.

\begin{figure}[h]
    \centering
    \includegraphics[width=0.48\textwidth]{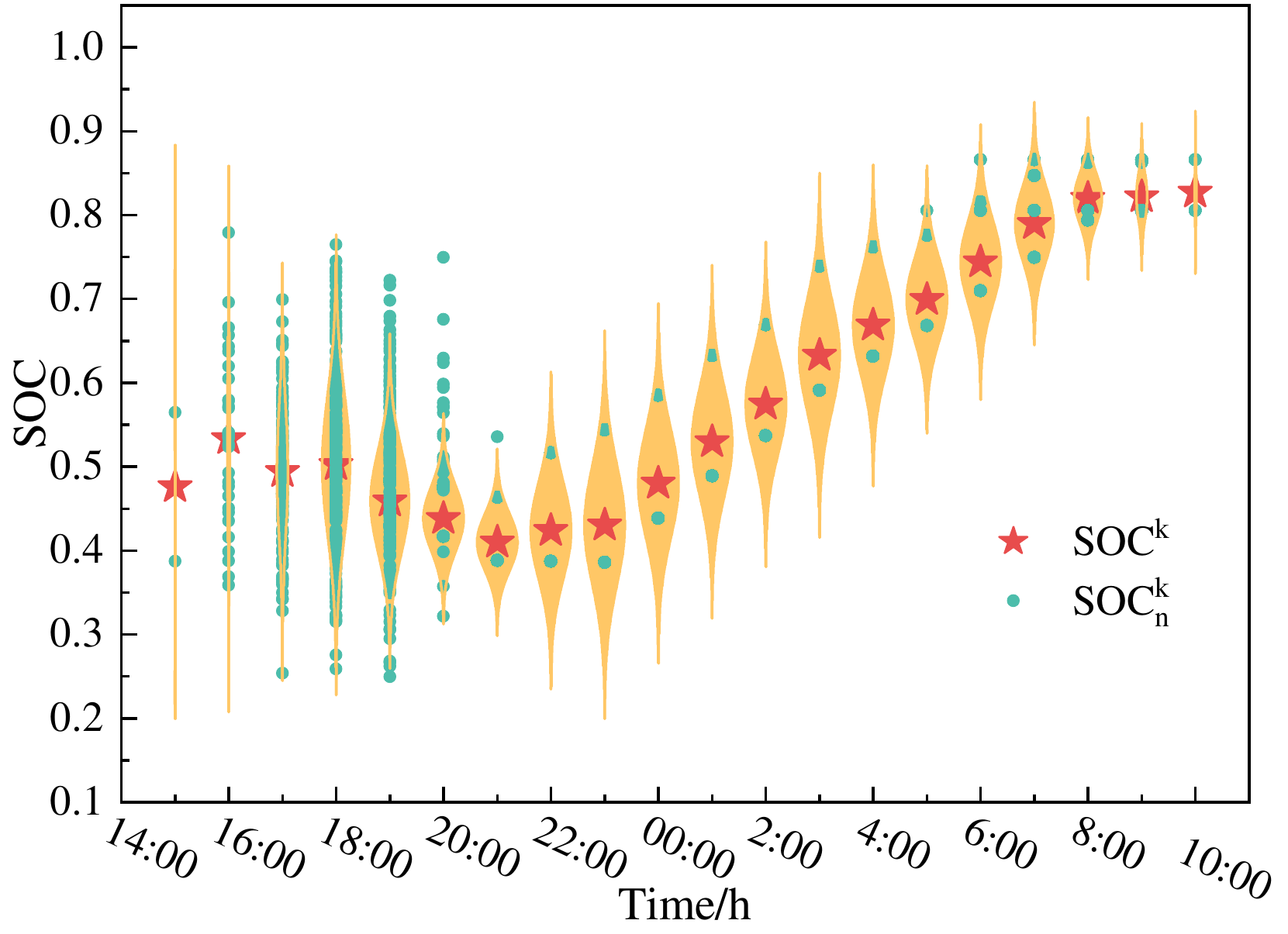}
		\caption{Distribution of individual EV SOC. }
		\label{fig:soc509pw}
\end{figure}

\begin{figure}[h]
    \centering
    \includegraphics[width=0.45\textwidth]{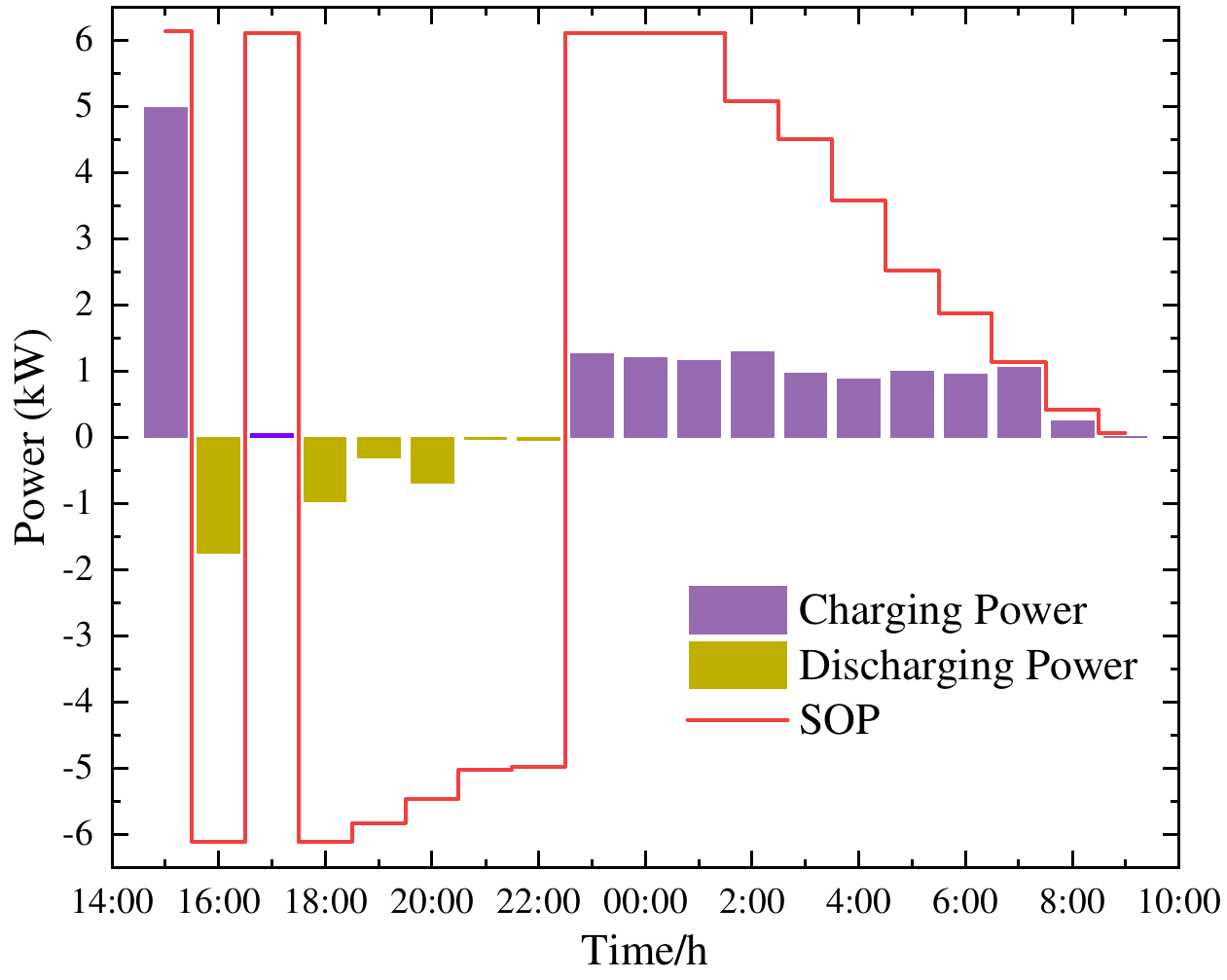}
    \caption{Single EV charging/discharging power in a day.}
    \label{fig:powersop}
\end{figure}

In Fig.~\ref{fig:soc509pw}, the red star indicates the SOC calculated based on the primary power strategy proposed by a selected validator in the context of the Proof of Stake algorithm. The scheduling period is from timeslot 15:00 to 10:00. At departure time, the SOC of EVA is 0.805, which meets the departure SOC boundary between 0.8 and 0.9. When EVA scheduling is completed, individual EV scheduling is required. The power allocation is elaborated in (\ref{kappa})-(\ref{soc1}). Therefore, the SOC of $n$-th EV should always be satisfied (\ref{deps}). The EV SOCs in the MACPO-based coordination scheduling time from timeslots 15:00 to 19:00 are more dispersed than in the remaining time because the EV SOCs differ greatly when the EVs arrive home for the charging/discharging schedule. The scheduling is based on the EVA SOC.  In the scheduling period, the SOC of the individual EV is always between 0.2 and 0.9, satisfying the SOC constraint of the individual EV. The bounded SOC constraints prevent battery overcharge and over-discharge, protect the battery from rapid degradation, and extend the battery's service life.
Taking an EV during the scheduling period as a case study, the relationship between power and SOP is elucidated. As demonstrated in Fig.~\ref{fig:powersop}, throughout the scheduling period, the charging/discharging power constantly satisfies the SOP constraints of the EV battery.

\subsection{Multi Stakeholder Benefits} 
MHVCS outperforms the five baseline strategies in terms of EI of the SOH, LV, and EV cost. Therefore, the proposed MHVCS can achieve multi-stakeholder benefits in the EVA coordination strategy. 

The three stakeholders — the EV user, EVA, and DSO — have different utility goals. The MHVCS can achieve nearly identical one-year SOHs and capacity degradation curves for EV users as the BL1 uncontrolled and BL2 optimal charging strategies. With the MACPO-based MHVCS coordination strategy, EV users participating in the charging/discharging coordination in the V2G service will not experience significant changes in battery capacity degradation compared to using only the charging service. There was no significant shift in charging costs due to the MHVCS charging/discharging control strategy. The SOC and SOP constraints are also satisfied during the coordination period, as shown in Fig.~\ref{fig:soc509pw} and Fig.~\ref{fig:powersop}. MHVCS can simultaneously meet EV travel needs and EVA energy demand.
The EVA incurs the smallest DSO cost, implying the highest revenue compared to the five baseline strategies. Furthermore, the DSO achieves the lowest LV to mitigate load fluctuation by integrating renewable energy.
MHVCS enables all three stakeholders to achieve their goals with mutual benefits.

\section{Conclusion}
This paper presents a MHVCS based on MACPO and PoS algorithms for the integration of renewable energy and the stability of the power grid. The proposed strategy considers the interests of the DSO, EVAs, and users and utilizes a hierarchical scheduling framework. On the DSO side, load fluctuations and DSO cost are considered, while
on the EVA side, energy constraints and charging costs are considered.
The three critical battery conditioning parameters of battery SOX are regarded on the
user side, including state of charge, state of power, and state of health. 
Moreover, the MACPO algorithm can ensure that the battery state constraints, energy constraints and other conditions are satisfied. Besides, our MHVCS strategy achieves good scheduling results for scheduling multiple electric vehicle aggregates.
The comparative analysis demonstrates the outstanding, flexible, adaptable, and scalable performance of the proposed hierarchical V2G coordination strategy for large-scale V2G under realistic operating conditions. 
Compared with the five baseline strategies, the proposed MHVCS achieves better performance in fattening the load variance. The proposed method also features less battery degradation and lower DSO cost.
The findings contribute to the research on renewable energy and V2G by providing a novel approach to V2G coordination that can optimize energy utilization, improve the stability of the power grid, and benefit all involved stakeholders. More research can be conducted to explore additional applications of multi-agent deep reinforcement learning in renewable energy integration and smart grid management.

\appendices

\section{Derivation of Multi-agent Constrained Policy Optimization}\label{appb}
\subsection{Single Agent Policy Optimization with Constraints}
Before proceeding to a multi-agent context, we illustrate the policy optimization method for single agent under constraint. Trust Region Policy Optimization\cite{schulman2015trust}, as one of the efficient policy-based DRL algorithms, guarantees monotonic improvement for general stochastic policies through the trust region method, considering the chances that the policy might deteriorate due to a large update step when the policy network is deep. According to the standard definitions of DRL, the state action value function is $Q_\pi(s_k,a_k)\doteq\mathbb{E}_{S_{k+1},A_{k+1},...}\left [\sum_{l=0}^{K-k}\gamma^lR_{k+l} \right]$, 
the state value function being 
$V_\pi(s_k)\doteq\mathbb{E}_{A_k,S_{k+1},A_{k+1},...}\left [\sum_{l=0}^{K-k}\gamma^lR_{k+l} \right ]$, 
and the advantage function is denoted as,
$A_\pi(s_k,a_k)\doteq Q_\pi(s_k,a_k)-V_\pi(s_k)$. 
For policy learning, the primal objective function is computed by the total expected rewards 
$J(\pi) =\mathbb{E}_{S_0}\left [V_\pi(S_0)\right ]=\mathbb{E}_{\tau \sim \pi}\left[\sum_{k=0}^{T} \gamma^{k} R_k\right]$. 
The improvement of the objective function from the old policy $\pi$ to another new policy $\widetilde{\pi}$ can be proven as the summation of advantage function over states.
\begin{align}
     &J(\widetilde{\pi})-J(\pi)\doteq \sum_{s}\rho_{\widetilde{\pi}}(s)\sum_{a}\widetilde{\pi}(a|s)A_\pi(s,a)
\end{align}
Since the new policy $\widetilde{\pi}$ is unknown in advance, it will be a huge job to try every $\widetilde{\pi}$ and collect trajectories to compute the Monte-Carlo Approximation of the state visitation density $\rho_{\widetilde{\pi}}$. Thus, as long as step $\pi\to\widetilde{\pi}$ is sufficiently small, a fairly accurate local approximation to $J\left (\widetilde{\pi}\right )$ is as follows, defined as $L_\pi(\widetilde{\pi})$
\begin{align}
    &L_\pi(\widetilde{\pi})
    =J(\pi) + \sum_{s}\rho_\pi(s)\sum_{a}\widetilde{\pi}(a|s)A_\pi(s,a) \approx J\left (\widetilde{\pi}\right )
\end{align}

Then, a lower bound of the objective function was theoretically justified, denoted by surrogate function $M_\pi\left ( \widetilde{\pi}\right )$
\begin{align} \label{eq1}
    J(\widetilde{\pi})\ge M_\pi(\widetilde{\pi})&\doteq L_\pi(\widetilde{\pi})-CD_{KL}^{max}(\pi,\widetilde{\pi}),\\
    where\ \  C&=\frac{4\epsilon \gamma }{(1-\gamma)^2}, \notag\\
    \epsilon &=\max_{s}\mathbb{E}_{a\sim\widetilde{\pi}}\left [A_\pi(s,a)\right ],\notag\\ 
    D_{KL}^{max}(\pi,\widetilde{\pi})&=\max_{s} D_{KL}(\pi,\widetilde{\pi})\notag
\end{align}

In equality~(\ref{eq1})  implies that any policy that improves $M$ is guaranteed to have the non-decreasing feature of $J\left ( \widetilde{\pi} \right ) $. By replacing the KL-penalty terms with trust region constraint, the policy network could take a larger but robust step. Hence, a policy update scheme was proposed as a solution to the following optimization problem. Note that in practice, a heuristic approximation to the maximum KL divergence is applied.
\begin{align}
    \pi_{k+1}&=\max_{\pi}L_{\pi_k}(\pi)\\
     s.t.\ &\overline{D_{KL}^{\rho}}  (\pi_{k+1},\pi)\le\delta \notag
\end{align}

In addition, to avoid the large memory space that could be occupied by the Hessian matrix, if updated directly through the closed form, the use of the conjugate gradient method is effective. 
Then adopt the linear search later in the event of violation of the KL-divergence constraint. 
The GAE and Monte Carlo are commonly employed to estimate the advantage.

Motivated by the trust region method and the monotonically improvement theorem, Achiam et al. proposed CPO for CMDP \cite{achiam2017constrained}. Defining an auxiliary cost function as $C:\mathcal{S}\times \mathcal{A}\to\mathcal{C}$ and the limit $d$. Similarly to expected rewards, the action-state value, state value, and advantage function with respect to cost are denoted as 
$Q_C^\pi\doteq \mathbb{E}_{S_{k+1},A_{k+1},...}\left [\sum_{l=0}^{K-k}\gamma^lC_{k+l} \right]$, 
$V_C^\pi\doteq\mathbb{E}_{A_k,S_{k+1},A_{k+1},...}\left [\sum_{l=0}^{K-k}\gamma^lC_{k+l} \right ]$, 
$A_C^\pi\doteq Q_C^\pi-V_C^\pi$. 
The expected total discounted cost function in analogy to $J(\pi)$ is $J_C(\pi)$, then a tight upper bound of $J_C(\widetilde{\pi})$ was proved during the update iteration. That is, for any policies $\widetilde{\pi},\pi$
\begin{align}\label{eq2}
    J_{C}(\widetilde{\pi})&\le J_C(\pi)+\frac{1}{1-\gamma} \mathbb{E}_{s\sim\rho_\pi,a\sim\widetilde{\pi}}\left [A_C^\pi(s,a)\right ] \notag\\
    &+\frac{\sqrt{2} \gamma\epsilon_C^{\widetilde{\pi}}}{1-\gamma }\sqrt{\mathbb{E}_{s\sim\rho_\pi}\left [ D_{KL}(\widetilde{\pi},\pi|s) \right ]}
\end{align}
where $\epsilon_C^{\widetilde{\pi}}\doteq \max_{x}\left [ A_C^\pi(s,a) \right ] $. Hence, as long as the updated polices $\widetilde{\pi}$ could restrict the right-hand side in equation (\ref{eq2}) to the limit $d$, then safety can be guaranteed, leading to constrained policy update scheme. 
\begin{align}\label{eq3}
      \pi_{k+1}&=arg\max_{\pi} \mathbb{E}_{s\sim\rho_{\pi_k},a\sim\pi}\left [A^{\pi_k}(s,a)\right]\\
        s.t.&\ J_C(\pi_k)+\frac{1}{1-\gamma }\mathbb{E}_{s\sim\rho_{\pi_k},a\sim\pi}\left [A_C^{\pi_k}(s,a)\right]\le c, \notag\\
        &D_{KL}(\pi_k,\pi_{k+1})\le\delta \notag
\end{align}

We note that the trust region method has already been used in the formulation in~\ref{eq3} to allow for a larger step rather than adding KL-penalty terms to both the objective function and the upper bound of the cost constraint.

\subsection{Multi-agent Trust Region Policy Optimization with Constraints}
Under the Multi-agent setting, we extend the definitions and notations above, with each policy parameterized by $\theta^i$. Denoting the joint policy as $ \pi(\cdot |s_k)=\prod_{i=1}^{N}\pi_i(\cdot^i|s_k) $, the joint action as $\boldsymbol{a}_k=\left ( a_k^1,a_k^2,...a_k^N \right )$. Then, the action-state value function, state value function and advantage function in terms of rewards are defined to be the following equations: 
\begin{align}
    &Q_{\pi}(s,\boldsymbol{a})\doteq \mathbb{E}_{\tau_{1:k}\sim\pi}\left [\sum_{k=0}^{k=K} \gamma ^tR_k(s_k,\boldsymbol{a_k}) |s_0=s,\boldsymbol{a_0}=\boldsymbol{a}\right ] \notag\\ 
    &V_{\pi}(s)\doteq \mathbb{E}_{\boldsymbol{a}\sim\pi,\ s_{1:k}\sim\rho _\pi}\left [ \sum_{k=0}^{k=K} \gamma ^kR_k(s_k,\boldsymbol{a_k}) |s_0=s\right ] \notag\\
    &A_{\pi}(s,\boldsymbol{a})\doteq Q_{\pi}(s,\boldsymbol{a})-V_{\pi}(s) \notag
\end{align}

Replacing rewards above with cost, assuming that the dimension of cost is equal to one in our model, we can get the cost counterpart individually: 
\begin{align}
    &Q_{C,\pi}^i(s,a^i)\doteq\mathbb{E}_{\boldsymbol{a^{-i}}\sim\pi^{-i},\ a^i_{1:k}\sim\pi^i,\ s_{1:k}\sim\rho _\pi } \notag\\
    &\quad\qquad\qquad\qquad \left[\sum_{k=0}^{k=K} \gamma ^kC_k^i(s_k,a_k^i) |s_0=s,a_0=a^i\right ] \notag\\ 
    &V_{C,\pi}^i(s)\doteq \mathbb{E}_{\boldsymbol{a}\sim\pi,s_{1:k}\sim\rho _\pi}\left [ \sum_{k=0}^{k=K} \gamma ^kC_k^i(s_k,a_k^i) |s_0=s\right ] \notag\\
    &A_{C,\pi}^i(s,a^i)\doteq Q_{C,\pi}^i(s,a^i)-V_{C,\pi}^i(s) \notag
\end{align}

In multi-agent CMDP, the aggregators are aimed at finding the optimal joint policy that maximizes the expected total reward 
$J(\pi)\doteq
\mathbb{E}_{\boldsymbol{\tau }\sim \pi}\left[\sum_{k=0}^{K} \gamma^{k} R\left(s_k, \boldsymbol{a}_{k}\right)\right]$ 
and keeping the expected total cost
$J_C^i(\pi)\doteq
\mathbb{E}_{\tau\sim\pi}\left[\sum_{k=0}^{k} \gamma^{k} C^i\left(s_k, a_k\right)\right]$
within the bound $d^i$ for any agent $i$.

Kuba et al. developed HATRPO to enable heterogeneous multiple agents, {\it{i.e.}}, no parameter sharing between agents' policy network, to learn monotonically improving policies \cite{kuba2022trust}. According to their work, they highlighted the performance of different groups from $\mathcal{N}$. Defining the action value function with respect to subset $i_{1:m}$, a random subset of $\mathcal{N}$ with the size of $m$, its complement denoted by $-i_{1:m}$.
\begin{align}
    Q^{i_{1:m}}_\pi (s, \boldsymbol{a}^{i_{1:m}})\doteq\mathbb{E}_{\boldsymbol{a}^{-i_{1:m}}\sim \pi^{-i_{1:m}}}\left [ Q_\pi\left (s,\boldsymbol{a}^{-i_{1:m}},\boldsymbol{a}^{i_{1:m}}\right )  \right ]
\end{align}
and the advantage function with regard to disjoint subset
\begin{align}
    A^{i_{1:m}}_\pi (s, \boldsymbol{a}^{i_{1:p}},\ \boldsymbol{a}^{i_{1:m}})&\doteq Q^{i_{1:p},i_{1:m}}_\pi (s, \boldsymbol{a}^{i_{1:p}},\boldsymbol{a}^{i_{1:m}})\notag\\
    &-Q^{i_{1:p}}_\pi (s, \boldsymbol{a}^{i_{1:p}})
\end{align}

Next, an important lemma, which serves as the transition from single agents to multi-agents, is the decomposition of the advantage function without any requirements on the decomposability of the joint advantage function. That is 
\begin{align}\label{eq4}
    A_\pi^{i_{1:m}}\left (s,\boldsymbol{a}^{i_{1:m}} \right ) =\sum_{j=1}^{m}A_\pi^{i_j}\left(s,\boldsymbol{a}^{i_{1:j-1}},a^{i_j}\right) 
\end{align}

Equation (\ref{eq4}) implies the positive assurance of $A_\pi$ as long as the agents take actions sequentially in an arbitrary order, while ensuring that the advantage over every action $A_\pi^{i_j}\left(s,\boldsymbol{a}^{i_{1:j-1}},a^{i_j}\right)$
is greater than zero, which is the fundamental idea behind HATRPO.

Building on (\ref{eq4}), the generalization from Equation (\ref{eq1}) to Multi-agent problem was derived
\begin{align}
    &J\left (\widetilde{\pi } \right )\ge 
\sum_{m=1}^{N} \left [ L_{\pi}^{i_{1:m}}\left (\widetilde{\pi} ^{i_{1:m-1}},\widetilde{\pi } ^{i_m}\right )
 - CD_{KL}^{max}\left ( \widetilde{\pi} ^{i_m},\pi^{i_m} \right) \right ] \\
     &where\qquad L_{\pi}^{i_{1:m}}\left (\widetilde{\pi} ^{i_{1:m-1}},\widetilde{\pi } ^{i_m}\right )\doteq \notag\\
     &\mathbb{E}_{s\sim\rho_\pi,\boldsymbol{a}^{i_{1:m-1}}\sim\widetilde{\pi}^{i_{1:m-1}}}\left [ A_\pi^{i_m}\left ( s,\boldsymbol{a}^{i_{1:m-1}}, a^{i_m}\right )  \right ]  \notag
\end{align}

This inequality indicates that if agents policies are optimized sequentially, we can guarantee an increment on the lower bound of objective function, further, the monotonic improvement of joint performance.

For the purpose of incorporating the cost constraints to the multi-agent problem above, extending the "surrogate" idea to approximate the performance of $\widetilde{\pi}$ on cost.
\begin{align}
    &L_C^i(\pi^i) \triangleq \mathbb{E}_{a^i\sim \pi^i}\left[ A^i_{C,\pi}\left ( s,a^i \right ) \right]
\end{align}

Using Equation \ref{eq2}, together with the multi-agent advantage decomposition property, a tight upper bound of $J_C^i\left(\widetilde{\pi}\right)$ was theoretically-justified
\begin{align}
    J_C^{i}(\widetilde{\pi}) &\leq J_C^{i}(\pi)+L_{C,\pi}^{i}\left(\widetilde{\pi}^{i}\right)+\nu^{i} \sum_{j=1}^{N}  D_{K L}^{\max }\left(\pi^{i}, \widetilde{\pi}^{i}\right) \notag\\
    &\text { where } \nu^{i}=\frac{4 \gamma \max _{s, a^{i}}\left|A_{C,\pi}^{i}\left(s, a^{i}\right)\right|}{(1-\gamma)^{2}}
\end{align}

This lemma implies that if the KL-divergence between two policies is small enough, the change in expected cost for each agent can be controlled by the surrogate cost, which provides guidance to update the joint policies with a guarantee of satisfaction of safety constraints monotonic, in addition to the improvement of reward performance, {\it{i.e.}} 
$J\left(\widetilde{\pi}\right)\ge J\left(\pi\right)$ as well as $J^i\left(\pi\right)\le c^i,\forall i\in \mathcal{N} $.
Therefore, if agents update their policies following a sequential update scheme, the new policy could improve the total return. Specifically, every agent solves the following optimisation problem sequentially:
\begin{align}
    &\widetilde{\pi}^i = \max L_\pi\left ( \widetilde{\pi}^i \right ) - \rho D_{KL}^{max}\left ( \widetilde{\pi}^i,\pi^i \right )\\
    &D_{KL}^{max} \doteq\max_{s} D_{KL}\left (\widetilde{\pi}^i\left ( \cdot |s \right ) ,\pi^i\left ( \cdot |s \right ) \right ) \notag
\end{align}

Implementing constraints as a penalty term added to the total reward function could result in a dilemma, that is, if $\rho$ is too small, the agent might overlook the constraint term, whereas the agent will stay still to satisfy the constraints, rendering slow performance improvement. Also, considering the difficulty in computing the maximum KL-penalty, one can still adopt the trust region method. Consequently, the problem is rewritten as follows.
\begin{align}
    &\widetilde{\theta^i}=\arg \max_{\theta}L(\pi) \\
    &s.t.\  J_C^i\left (\pi\right )+ L^i(\pi^i) \le c^i \notag\\
    &\qquad \mathbb{E}\left [ D_{KL}\left (\widetilde{\pi}^i,\pi^i \right) \right]\le c^i\notag
\end{align}
Through Taylor expansion, the objective function and cost constraint could be approximated linearly or quadratically for the KL-divergence. 
\begin{align}
    &\theta_{k+1}=arg\max_{\theta }\ g^T(\theta -\theta_k)\\
    &s.t.\ c+b^T(\theta -\theta_k)\le 0\notag\\
    &\qquad\frac{1}{2} (\theta-\theta_k)^TH(\theta-\theta _k)\le \theta\notag
\end{align}

Lastly, to deal with the issue of the approximation error, the backtracking line search succeeded in guaranteeing constraint satisfaction. When it comes to infeasible cases, a limiting search enables the policy to recover from violation, {\it{i.e.}}, set the update direction to the decrease of the constraint value.
\begin{align}
    \theta_{k+1}=\theta_k-\sqrt{\frac{2\delta}{b^TH^{-1}b}}H^{-1}b 
\end{align}

\section{Simulation Environment with Battery Conditioning}\label{sebc}
Battery conditioning is a process that involves maintaining and improving battery performance through a series of controlled charging/discharging cycles. It extends the battery's lifespan, improves capacity retention, and ensures optimal performance under varying operating conditions. Taking into account the actual state of the battery, we propose battery states (including SOH, SOP, and SOC) constraints in the coordination control strategy.
\subsubsection{SOH Model}
Saxena et al.\cite{saxena2016cycle} established a capacity attenuation model for batteries under full or partial cycle conditions and obtained the relationship between average SOC and SOC variation and the capacity loss rate. SOH is shown as:

\begin{align}
&SOH^w_n(\%)=100 - 3.25{SOC_{n,ave}^w}(1+3.25\Delta SOC_n^w\nonumber\\
&\qquad \qquad \qquad-2.25{\Delta SOC_n^w}^2)*(w/100)^{0.453}\\
&{SOC_{n,ave}^w}=0.5(\overline{SOC_n^w}+\underline{SOC_n^w})\\
&\Delta SOC_n^w=\overline{SOC_n^w}-\underline{SOC_n^w}\\
&w(m)=w(m-1)+\varepsilon (m-1) M_1\\
&\varepsilon (m)=\frac{0.5}{M(m-1)} (2-\frac{DOD(m-2)+DOD(m)}{DOD(m-1)})\nonumber\\
&\qquad \quad +\varepsilon (m-1)\\
&M(m)=H{{(\frac{DOD(m)}{100})}^{-\kappa }} {{I}_{dis,ave}}{{(m)}^{-{{\gamma }_{1}}}} {{I}_{ch,ave}}{{(m)}^{-{{\gamma }_{2}}}}
\end{align}
where $SOC_{n,ave}^w$ denotes the average SOC during the $w-{th}$ equivalent full cycle of the $n$-th EV. $\Delta SOC_n^w$ represents the variation in SOC during the $w-{th}$ equivalent full cycle of the $n$-th EV. Furthermore, $w$ indicates the equivalent complete cycles endured by the battery. $\varepsilon$ is the battery aging factor, $M_1$ is the equivalent full cycle constant, and is set to 1500 in the model. $M$ is the maximum number of cycles. $DOD$ stands for depth of discharge which refers to the percentage of the battery capacity that has been discharged. $H$ is the cycle number, $\kappa$ is the exponent factor for the DOD, ${I}_{{dis,ave}/{ch,ave}}$ is the average discharge/charge current during a half-cycle duration. ${\gamma }_{1}$ and ${\gamma }_{2}$ are the exponent factors for the discharge / charge current, respectively. 

\subsubsection{SOP Model}
The battery SOP refers to the maximum charging/discharging power that the battery can withstand within a certain period of time. It is important to avoid overcharging and over-discharging of the battery and to extend its useful life. In the model, the effects of SOC, battery terminal voltage, and battery design are considered.

To prevent the SOC from exceeding the limit due to battery charging or discharging, the charging/discharging current of the battery is constrained on the basis of the current SOC. The maximum charging/discharging current of the battery is set during the $k$ to $k+n$ time periods.
\begin{align}
&\overline{I^{k}_{SOC,dis}}=\cdot \cdot \cdot =\overline{I^{k+n}_{SOC,dis}}=\frac{Q_n(SOC^k-\underline{SOC})}{n{{k}_{s}}}\\
&\overline{I^{k}_{SOC,ch}}=\cdot \cdot \cdot =\overline{I^{k+n}_{SOC,ch}}=\frac{Q_n(\overline{SOC}-SOC^k)}{n{{k}_{s}}}
\end{align}
where, $\overline{I^{k}_{SOC,dis}}$ is the maximum instantaneous discharge current based on battery SOC constraints at timeslot $k$. $\overline{I^{k}_{SOC,ch}}$ is the maximum instantaneous charging current based on the SOC constraints of the battery at the time range $k$. ${{k}_{s}}$ is the sampling interval. ${{Q}_{n}}$ is the maximum current available capacity of the lithium battery.

The battery voltage at the terminal cannot exceed the design limit during operation, so the battery terminal voltage is constrained. The maximum charging/discharging current of the battery is set during $k$ to $k+n$ time periods.
\begin{align}
&\overline{I^{k}_{{U_T},dis}}=\cdot \cdot \cdot =\overline{I^{k+n}_{{U_T},dis}}=\frac{{{U}_{oc}}(SOC^{k})-\underline{U_{k}}}{{{R}_{0}}+\frac{n{{k}_{s}}}{{{Q}_{n}}}* \frac{\partial {U_{OC}}(SOC_{k})}{\partial SOC_{k}}}\\
&\overline{I^{k}_{{U_T},ch}}=\cdot \cdot \cdot =\overline{I^{k+n}_{{U_T},ch}}=\left| \frac{{{U}_{oc}}(SOC^k)-\overline{U_T}}{R_0+\frac{n{T_s}}{Q_n}* \frac{\partial {U_{OC}}(SOC^k)}{\partial SOC^k}} \right|
\end{align}
where $\overline{I^{k}_{{U_T},dis}}$ is the maximum instantaneous discharge current based on battery terminal voltage constraints at timeslot $k$.
$\overline{I^{k}_{{U_T},ch}}$ is the maximum instantaneous charging current based on the voltage constraints of the battery terminal at the time slot $k$.
${U_{oc}}(SOC^k)$ is the battery terminal voltage at timeslot $k$.
${R_0}$ is the internal resistance of the battery.
$\overline{U_T}$ and $\underline{U_T}$ are the limits of the battery terminal voltage.

The peak current of the lithium battery under multiple constrained factors.
\begin{align}
&\overline{I^{k}_{dis}}=\min \{ \overline{I^{k}_{SOC,dis}},\overline{I^{k}_{{U_T},dis}},\overline{I^{k}_{{design},dis}}\}\\
&\overline{I^{k}_{ch}}=\min \{ \overline{I^{k}_{SOC,ch}},\overline{I^{k}_{{U_T},ch}},\overline{I^{k}_{{design},ch}}\}
\end{align}
where $\overline{I^{k}_{dis}}$ is the maximum discharge current of the battery at timeslot $k$.
$\overline{I^{k}_{ch}}$ is the maximum charging current of the battery at timeslot $k$.
$\overline{I^{k}_{{design},dis}}$ is the maximum discharge current of the battery design.
$\overline{I^{k}_{{design},ch}}$ is the maximum charging current of the battery design.

The $\overline{I^{k}_{dis}}$ and $\overline{I^{k}_{ch}}$ can be used to obtain the continuous peak charging/discharging power of the battery.
\begin{align}
&\overline{P_{k}^{dis}}={U_T^k}* \overline{I^{k}_{dis}}\\
&\overline{P_{k}^{ch}}={U_T^k}* \overline{I^{k}_{ch}}
\end{align}
where $\overline{P_{k}^{dis}}$ is the continuous peak discharge power of the battery under multiple constraints at timeslot $k$.
$\overline{P_{k}^{ch}}$ is the continuous peak charging power of the battery under multiple constraints at timeslot $k$.
${U_T^k}$ is the battery terminal voltage at timeslot $k$.

\ifCLASSOPTIONcaptionsoff
  \newpage
\fi

\bibliographystyle{IEEEtran}
\bibliography{ref}

\end{document}